\documentclass[aps,prd,twocolumn,floatfix,10pt,amssymb,amsfont,amsmath,groupaddress,float,nofootinbib,longbibliography]{revtex4-2}

\usepackage{preamb}

\pgfdeclarelayer{back}
\pgfdeclarelayer{middle}
\pgfdeclarelayer{front}
\pgfsetlayers{back,middle,main,front}

\tikzset{->1-/.style={decoration={
			markings,
			mark=at position #1 with {\arrow{Triangle[fill=black, width=3.2pt, length=3.2pt]}}},postaction={decorate}}}
			
\tikzset{
	every picture/.append style={
		line join=round,
		line cap=round,
		line width = 0.6pt
	}
}

\tikzset{mask point/.style={ transform shape, sloped, 
minimum width=20pt, minimum height=4pt, inner sep=0pt, ultra thin, font=\tiny}}

\tikzset{
    clip even odd rule/.code={\pgfseteorule},
    invclip/.style={clip,insert path=[clip even odd rule]{
    [reset cm](-\maxdimen,-\maxdimen)rectangle(\maxdimen,\maxdimen)
    }}}

\tikzstyle{wig}=[color=BrickRed, decorate,decoration={snake, amplitude=0.6pt, segment length=3.8pt}, line width=1.1pt]
\tikzstyle{wigb}=[color=BlueViolet, decorate,decoration={snake, amplitude=0.6pt, segment length=3.8pt}, line width=1.1pt]
\tikzstyle{dot}=[line width=1pt, line cap=round, dash pattern=on 0pt off 2\pgflinewidth]

\newcommand{\lattice}[0]{
	\begin{tikzpicture}[scale=1.1,baseline={([yshift=-.5ex]current bounding box.center)},z={(-0.4,0.4)}]
	    \coordinate (a) at (3.5,0,1);
	    \draw[->,>={Triangle[fill=black, width=3pt, length=3pt]}] (a) -- ($(a)+(0.5,0,0)$);
	    \draw[->,>={Triangle[fill=black, width=3pt, length=3pt]}] (a) -- ($(a)+(0,0.5,0)$);
	    \draw[->,>={Triangle[fill=black, width=3pt, length=3pt]}] (a) -- ($(a)+(0,0,-0.5)$);
	    \node at ($(a)+(0.65,0,0)$) {${\sss \hat x}$};
	    \node at ($(a)+(0,0.7,0)$) {${\sss \hat y}$};
	    \node at ($(a)+(0,0,-0.7)$) {${\sss \hat z}$};
	    
	    \foreach \x in {0,1,2}{
	        \foreach \y in {0,1,2}{
	            \draw[color=black!50] (\x,\y,0) -- (\x,\y,1.5);
	        }
        }
	    \foreach \z/\col in {1/black!50,0/black}{
    	    \foreach \x in {0,1,2}{
                \draw[color=\col] (\x,-0.5,\z) -- (\x,1,\z);
    	        \draw[color=\col] (\x,1,\z) -- (\x,2.5,\z);
    	        \draw[color=\col] (-0.5,\x,\z) -- (1,\x,\z);
    	        \draw[color=\col] (1,\x,\z) -- (2.5,\x,\z);
    	    }
	    }
	    \node[yshift=-5pt, xshift=-5pt] at (0,0,0) {${\sss \msf v}$};
	    \node[yshift=-5pt] at (.5,0,0) {${\sss \msf e}$};
	    \node at (.5,.5,0) {${\sss \msf p}$};
	    \node[text=black!50] at (.5,.5,.5) {${\sss \msf c}$};
	\end{tikzpicture}
}

\newcommand{\cGate}[1]{
	\begin{tikzpicture}[scale=1.1,baseline={([yshift=-.5ex]current bounding box.center)},z={(-0.4,0.4)}]
        \draw (-1,0,0) -- (1,0,0) -- (1,1,0) -- (0,1,0) -- (0,0,0);
        \ifthenelse{#1=1}{
            \draw[dot, shorten <= 1pt, shorten >= 5pt] (-.5,0,0) -- (.5,.5,0);
            \node[fill=white,circle,inner sep=0pt] at (-.5,0,0) {${\sss {\rm c}}$};
            \node at (.5,.5,0) {${\sss \Sigma_X}$};
        }{}
        \ifthenelse{#1=2}{
            \node[fill=white,circle,inner sep=0pt] at (-.5,0,0) {${\sss g}$};
            \node at (.5,.5,0) {${\sss h}$};
        }{}
        \ifthenelse{#1=3}{
            \node[fill=white,circle,inner sep=0pt] at (-.5,0,0) {${\sss g}$};
            \node at (.5,.5,0) {${\sss [g+h]}$};
        }{}
        \ifthenelse{#1=4}{
            \draw[dot, shorten <= 1pt, shorten >= 5pt] (-.5,0,0) -- (.5,.5,0);
            \node[fill=white,circle,inner sep=0pt] at (-.5,0,0) {${\sss {\rm c}}$};
            \node at (.5,.5,0) {${\sss \Sigma_Z}$};
        }{}
        \ifthenelse{#1=5}{
            \node[fill=white,circle,inner sep=0pt] at (-.5,0,0) {${\sss g}$};
            \node at (.5,.5,0) {${\sss \omega^{h^g}}$};
        }{}
	\end{tikzpicture}
}

\newcommand{\cGateRelations}[1]{
	\begin{tikzpicture}[scale=1.1,baseline={([yshift=-.5ex]current bounding box.center)},z={(-0.4,0.4)}]
        \draw (0,0,0) -- (0,2,0) -- (1,2,0) -- (1,0,0);
        \draw (0,1,0) -- (1,1,0);
        \ifthenelse{#1=1}{
            \draw[dot, shorten <= 1pt, shorten >= 4pt] (0,.5,0) -- (.5,1.5,0);
            \node[fill=white,circle,inner sep=0pt] at (0,.5,0) {${\sss {\rm c}}$};
            \node at (.5,1.5,0) {${\sss \Sigma_Z^k}$};
            \draw[wig] (0,0,0) -- (1,0,0);
        }{}
        \ifthenelse{#1=2}{
            \draw[dot, shorten <= 1pt, shorten >= 3.5pt] (.5,1,0) -- (.5,1.5,0);
            \draw[dot, shorten <= 1pt, shorten >= 6pt] (1,.5,0) -- (.5,1.5,0);
            \node[fill=white,circle,inner sep=0pt] at (.5,1,0) {${\sss {\rm c}}$};
            \node[fill=white,circle,inner sep=0pt] at (1,.5,0) {${\sss {\rm c}}$};
            \node at (.5,1.5,0) {${\sss \Sigma_Z^k}$};
            \draw[wig] (0,0,0) -- (1,0,0);
        }{}
        \ifthenelse{#1=3}{
            \draw[dot, shorten <= 1pt, shorten >= 4pt] (0,.5,0) -- (.5,1.5,0);
            \node[fill=white,circle,inner sep=0pt] at (0,.5,0) {${\sss {\rm c}}$};
            \node at (.5,1.5,0) {${\sss \Sigma_Z}$};
            \draw[wigb] (0,0,0) -- (1,0,0);
        }{}
        \ifthenelse{#1=4}{
            \draw[dot, shorten <= 1pt, shorten >= 3.5pt] (.5,1,0) -- (.5,1.5,0);
            \draw[dot, shorten <= 1pt, shorten >= 6pt] (1,.5,0) -- (.5,1.5,0);
            \node[fill=white,circle,inner sep=0pt] at (.5,1,0) {${\sss {\rm c}}$};
            \node[fill=white,circle,inner sep=0pt] at (1,.5,0) {${\sss {\rm c}}$};
            \node at (.5,1.5,0) {${\sss \Sigma_Z^\dagger}$};
            \draw[wigb] (0,0,0) -- (1,0,0);
        }{}
	\end{tikzpicture}
}

\newcommand{\cGateCommute}[1]{
	\begin{tikzpicture}[scale=1.1,baseline={([yshift=-.5ex]current bounding box.center)},z={(-0.4,0.4)}]
        \draw (0,0,0) -- (0,2,0) -- (2,2,0) -- (2,0,0);
        \draw (1,0,0) -- (1,2,0);
        \draw (0,1,0) -- (2,1,0);
        \draw[wig] (0,0,0) -- (2,0,0);
        \ifthenelse{#1=1}{
            \draw[dot, shorten <= 1pt, shorten >= 3.5pt] (.5,1,0) -- (.5,1.5,0);
            \draw[dot, shorten <= 1pt, shorten >= 6pt] (1,.5,0) -- (.5,1.5,0);
            \draw[dot, shorten <= 1pt, shorten >= 4pt] (1,.5,0) -- (1.5,1.5,0);
            \node[fill=white,circle,inner sep=0pt] at (.5,1,0) {${\sss {\rm c}}$};
            \node[fill=white,circle,inner sep=0pt] at (1,.5,0) {${\sss {\rm c}}$};
            \node at (.5,1.5,0) {${\sss \Sigma_Z^k}$};
            \node at (1.5,1.5,0) {${\sss \Sigma_Z^k}$};
        }{}
        \ifthenelse{#1=2}{
            \draw[dot, shorten <= 1pt, shorten >= 3pt] (.5,1,0) -- (.5,1.5,0);
            \node[fill=white,circle,inner sep=0pt] at (.5,1,0) {${\sss {\rm c}}$};
            \node at (.5,1.5,0) {${\sss \Sigma_X}$};
            \node at (1.5,1.5,0) {${\sss \Sigma_X^\dagger}$};
        }{}
        \ifthenelse{#1=3}{
            \draw (0,2,0) -- (0,3,0) -- (2,3,0) -- (2,2,0);
            \draw (1,2,0) -- (1,3,0);
            \draw[dot, shorten <= 1pt, shorten >= 4pt] (1,1.5,0) -- (1.5,2.5,0);
            \draw[dot, shorten <= 1pt, shorten >= 4pt] (1,.5,0) -- (1.5,1.5,0);
            \draw[dot, shorten <= 1pt, shorten >= 4pt] (1,.5,0) -- (1.5,2.5,0);
            \node[fill=white,circle,inner sep=0pt] at (1,.5,0) {${\sss {\rm c}}$};
            \node[fill=white,circle,inner sep=0pt] at (1,1.5,0) {${\sss {\rm c}}$};
            \node at (1.5,2.5,0) {${\sss \Sigma_Z^k}$};
            \node at (1.5,1.5,0) {${\sss \Sigma_Z^k}$};
        }{}
        \ifthenelse{#1=4}{
            \draw (0,2,0) -- (0,3,0) -- (2,3,0) -- (2,2,0);
            \draw (1,2,0) -- (1,3,0);
            \node at (1.5,1.5,0) {${\sss \Gamma}$};
            \foreach \x/\y in {.5/1,1.5/1,1/.5,1/1.5}{
                \node[fill=white,rectangle,inner sep=1pt] at (\x,\y,0) {${\sss \sigma_X}$};
            }
        }{}
	\end{tikzpicture}
}

\newcommand{\hamOp}[1]{
	\begin{tikzpicture}[scale=1.1,baseline={([yshift=-.5ex]current bounding box.center)},z={(-0.4,0.4)}]
        \ifthenelse{#1=0}{
            \draw (0,0,0) -- (1,0,0);
            \node[fill=white,rectangle,inner sep=1pt] at (.5,0,0) {${\sss \sigma_Z}$};
        }{}
        \ifthenelse{#1=1}{
            \draw (0,0,0) rectangle (1,1,0);
            \foreach \x/\y in {0.5/0,0/0.5,0.5/1,1/0.5}{
                \node[fill=white,rectangle,inner sep=1pt] at (\x,\y,0) {${\sss \sigma_Z}$};
            }
        }{}
        \ifthenelse{#1=2}{
            \draw[color=black!50] (1,0,0) -- (1,0,1) -- (0,0,1) -- (0,0,0);
            \draw (0,0,0) -- (1,0,0);
            \foreach \x/\z in {0/0.5,0.5/1,1/0.5}{
                \node[fill=white,rectangle,inner sep=1pt,text=black!50] at (\x,0,\z) {${\sss \sigma_Z}$};
            }
            \node[fill=white,rectangle,inner sep=1pt] at (.5,0,0) {${\sss \sigma_Z}$};
        }{}
        \ifthenelse{#1=3}{
            \draw[color=black!50] (0,1,0) -- (0,1,1) -- (0,0,1) -- (0,0,0);
            \draw (0,0,0) -- (0,1,0);
            \node[fill=white,rectangle,inner sep=1pt] at (0,0.5,0) {${\sss \sigma_Z}$};
            \node[fill=white,rectangle,inner sep=1pt,text=black!50] at (0,0.5,1) {${\sss \sigma_Z}$};
            \node[fill=white,rectangle,inner sep=1pt,text=black!50] at (0,1,0.45) {${\sss \sigma_Z}$};
            \node[fill=white,rectangle,inner sep=1pt,text=black!50] at (0,0,0.55) {${\sss \sigma_Z}$};
        }{}
        \ifthenelse{#1=4}{
            \draw[color=black!50] (0,0,1) -- (0,0,0);
            \draw (-1,0,0) -- (1,0,0);
            \draw (0,-1,0) -- (0,1,0);
            \draw (1,0,0) -- (1,1,0) -- (0,1,0);
            \foreach \x/\y in {0.5/0,-0.5/0,0/0.5,0/-0.5}{
                \node[fill=white,rectangle,inner sep=1pt] at (\x,\y,0) {${\sss \sigma_X}$};
            }
            \node[fill=white,rectangle,inner sep=1pt,text=black!50] at (0,0,0.55) {${\sss \sigma_X}$};
            \node at (0.5,0.5,0) {${\sss \Gamma}$};
        }{}
        \ifthenelse{#1=5}{
            \draw[color=black!50] (0,0,0) -- (0,0,1) -- (0,1,1) -- (0,1,0);
            \draw (-1,0,0) rectangle (1,1,0);
            \draw (0,0,0) -- (0,1,0);
            \node at (-0.5,0.5,0) {${\sss \Sigma_X}$};
            \node at (0.5,0.5,0) {${\sss \Sigma_X^\dagger}$};
            \node[text=black!50] at (0,0.5,0.5) {${\sss \Sigma_X}$};
            \draw[dot, shorten <= 1pt, shorten >= 3pt] (-0.5,0,0) -- (-0.5,0.5,0);
            \node[fill=white,circle,inner sep=0pt] at (-0.5,0,0) {${\sss {\rm c}}$};
            \draw[dot,black!50, shorten <= 1pt, shorten >= 3pt] (0,0,0.5) -- (0,0.5,0.5);
            \node[fill=white,circle,inner sep=0pt,text=black!50] at (0,0,0.5) {${\sss {\rm c}}$};
        }{}
        \ifthenelse{#1=6}{
            \draw[color=black!50] (0,0,0) -- (0,0,1) -- (1,0,1) -- (1,0,0);
            \draw[dot,black!50, shorten <= 1pt, shorten >= 5pt] (0,0,0.5) -- (0.5,0,0.5);
            \node[fill=white,circle,inner sep=0pt,text=black!50] at (0,0,0.5) {${\sss {\rm c}}$};
            \draw (0,-1,0) rectangle (1,1,0);
            \draw (0,0,0) -- (1,0,0);
            \node at (0.5,0.5,0) {${\sss \Sigma_X}$};
            \node at (0.5,-0.5,0) {${\sss \Sigma_X^\dagger}$};
            \node[text=black!50] at (0.5,0,0.5) {${\sss \Sigma_X}$};
            \draw[dot, shorten <= 1pt, shorten >= 5pt] (0,-0.5,0) -- (0.5,-0.5,0);
            \node[fill=white,circle,inner sep=0pt] at (0,-0.5,0) {${\sss {\rm c}}$};
        }{}
        \ifthenelse{#1=7}{
            \draw[color=black!50] (0,0,0) -- (0,0,1) -- (0,1,1) -- (0,1,0);
            \draw (-1,0,0) rectangle (1,1,0);
            \draw (0,0,0) -- (0,1,0);
            \node at (-0.5,0.5,0) {${\sss \Sigma_X^\dagger}$};
            \node at (0.5,0.5,0) {${\sss \Sigma_X}$};
            \node[text=black!50] at (0,0.5,0.5) {${\sss \Sigma_X^\dagger}$};
            \draw[dot, shorten <= 1pt, shorten >= 5pt] (-0.5,0,0) -- (-0.5,0.5,0);
            \node[fill=white,circle,inner sep=0pt] at (-0.5,0,0) {${\sss {\rm c}}$};
            \draw[dot,black!50, shorten <= 1pt, shorten >= 5pt] (0,0,0.5) -- (0,0.5,0.5);
            \node[fill=white,circle,inner sep=0pt,text=black!50] at (0,0,0.5) {${\sss {\rm c}}$};
        }{}
        \ifthenelse{#1=8}{
            \draw[color=black!50] (0,0,0) -- (0,0,1) -- (1,0,1) -- (1,0,0);
            \draw[dot,black!50, shorten <= 1pt, shorten >= 5pt] (0,0,0.5) -- (0.5,0,0.5);
            \node[fill=white,circle,inner sep=0pt,text=black!50] at (0,0,0.5) {${\sss {\rm c}}$};
            \draw (0,-1,0) rectangle (1,1,0);
            \draw (0,0,0) -- (1,0,0);
            \node at (0.5,0.5,0) {${\sss \Sigma_X^\dagger}$};
            \node at (0.5,-0.5,0) {${\sss \Sigma_X}$};
            \node[text=black!50] at (0.5,0,0.5) {${\sss \Sigma_X^\dagger}$};
            \draw[dot, shorten <= 1pt, shorten >= 5pt] (0,-0.5,0) -- (0.5,-0.5,0);
            \node[fill=white,circle,inner sep=0pt] at (0,-0.5,0) {${\sss {\rm c}}$};
        }{}
        \ifthenelse{#1=9}{
            \draw[color=black!50] (0,0,0) -- (0,0,1) -- (1,0,1) -- (1,0,0);
            \draw[color=black!50] (0,0,1) -- (0,1,1);
            \draw[color=black!50] (1,0,1) -- (1,1,1);
            \draw[color=black!50] (0,1,0) -- (0,1,1) -- (1,1,1) -- (1,1,0);
            \node[text=black!50] at (.5,.5,1) {${\sss \Sigma_Z^\dagger}$};
            \node[text=black!50] at (.5,0,.5) {${\sss \Sigma^\dagger_Z}$};
            \node[text=black!50] at (1,.5,.5) {${\sss \Sigma^\dagger_Z}$};
            \node[text=black!50] at (.5,1,.5) {${\sss \Sigma_Z}$};
            \node[text=black!50] at (0,.5,.5) {${\sss \Sigma_Z}$};
            \draw[dot,black!50, shorten <= 1pt, shorten >= 5pt] (0,.5,1) to[out=65,in=180,looseness=1] (.5,1,.5);
            \node[fill=white,circle,inner sep=0pt,text=black!50] at (0,.5,1) {${\sss {\rm c}}$};
            \draw[dot,color=black!50, shorten <= 1pt, shorten >= 5pt] (.5,0,1) to[out=65,in=180,looseness=1] (1,.5,.5);
            \node[fill=white,circle,inner sep=0pt,text=black!50] at (.5,0,1) {${\sss {\rm c}}$};
            \draw[dot, shorten <= 1pt] (0,0,.5) to[out=0,in=-115,looseness=1] (.33,.51,0);
            \node[fill=white,circle,inner sep=0pt,text=black!50] at (0,0,.5) {${\sss {\rm c}}$};
            \node at (.5,.5,0) {${\sss \Sigma_Z}$};
            \draw (0,0,0) rectangle (1,1,0);
        }{}
        \ifthenelse{#1=10}{
            \draw[color=black!50] (0,0,0) -- (0,0,1) -- (1,0,1) -- (1,0,0);
            \draw[color=black!50] (0,0,1) -- (0,1,1);
            \draw[color=black!50] (1,0,1) -- (1,1,1);
            \draw[color=black!50] (0,1,0) -- (0,1,1) -- (1,1,1) -- (1,1,0);
            \node[text=black!50] at (0.5,.5,1) {${\sss \Sigma_Z}$};
            \node[text=black!50] at (.5,0,.5) {${\sss \Sigma_Z}$};
            \node[text=black!50] at (1,.5,.5) {${\sss \Sigma_Z}$};
            \node[text=black!50] at (.5,1,.5) {${\sss \Sigma^\dagger_Z}$};
            \node[text=black!50] at (0,.5,.5) {${\sss \Sigma^\dagger_Z}$};
            \draw[dot,black!50, shorten <= 1pt, shorten >= 5pt] (0,.5,1) to[out=65,in=180,looseness=1] (.5,1,.5);
            \node[fill=white,circle,inner sep=0pt,text=black!50] at (0,.5,1) {${\sss {\rm c}}$};
            \draw[dot,color=black!50, shorten <= 1pt, shorten >= 5pt] (.5,0,1) to[out=65,in=180,looseness=1] (1,.5,.5);
            \node[fill=white,circle,inner sep=0pt,text=black!50] at (.5,0,1) {${\sss {\rm c}}$};
            \draw[dot, shorten <= 1pt] (0,0,.5) to[out=0,in=-115,looseness=1] (.33,.51,0);
            \node[fill=white,circle,inner sep=0pt,text=black!50] at (0,0,.5) {${\sss {\rm c}}$};
            \node at (.5,.5,0) {${\sss \Sigma^\dagger_Z}$};
            \draw (0,0,0) rectangle (1,1,0);
        }{}
	\end{tikzpicture}
}

\newcommand{\ZeroCharge}[0]{
	\begin{tikzpicture}[scale=1.1,baseline={([yshift=-.5ex]current bounding box.center)},z={(-0.4,0.4)}]
	    
	    \foreach \x in {0,1,2}{
	        \foreach \y in {0,1,2}{
	            \draw[color=black!50] (\x,\y,0) -- (\x,\y,1.5);
	        }
        }
	    \foreach \z/\col in {1/black!50,0/black}{
    	    \foreach \y in {0,1,2}{
                \draw[\col] (-.5,\y,\z) -- (2.5,\y,\z);
    	    }
    	    \foreach \x in {0,1,2}{
                \draw[\col] (\x,-.5,\z) -- (\x,2.5,\z);
    	    }
	    }
	    \foreach \x/\y in {.5/0,1/.55,1.5/1,2/1.55}{
            \node[fill=white,rectangle,inner sep=1pt] at (\x,\y,0) {${\sss \sigma_Z}$};
	    }
	    \node[fill=BrickRed,circle,inner sep=1.5pt] at (0,0,0) {};
	    \node[fill=BrickRed,circle,inner sep=1.5pt] at (2,2,0) {};
	    \node[yshift=-5pt, xshift=-5pt,text=BrickRed] at (0,0,0) {${\sss e}$};
	    \node[yshift=5pt, xshift=5pt,text=BrickRed] at (2,2,0) {${\sss e}$};
	\end{tikzpicture}
}

\newcommand{\wigLine}[1]{
	\begin{tikzpicture}[scale=1.1,baseline={([yshift=-.5ex]current bounding box.center)},z={(-0.4,0.4)}] 
	    \ifthenelse{#1=1}{
	        \draw[wig] (0,0,0) -- (.5,0,0);
	    }{}
	    \ifthenelse{#1=2}{
	        \draw[wigb] (0,0,0) -- (.5,0,0);
	    }{}
	\end{tikzpicture}
}

\newcommand{\condString}[0]{
	\begin{tikzpicture}[scale=1.1,baseline={([yshift=-.5ex]current bounding box.center)},z={(-0.4,0.4)}]
	    
	    \fill[BlueViolet, fill opacity=0.15] (2,1,0) -- (3,1,0) -- (3,1,1) -- (2,1,1) -- (2,1,0);
	    \fill[BlueViolet, fill opacity=0.15] (2,0,0) rectangle (3,2,0);
	    	    
	    \foreach \x in {0,1,2,3,4,5}{
	        \foreach \y in {0,1,2}{
	            \draw[color=black!50] (\x,\y,0) -- (\x,\y,1.5);
	        }
        }
	    \foreach \z/\col in {1/black!50,0/black}{
    	    \foreach \y in {0,1,2}{
    	        \ifthenelse{\equal{\y}{1} \AND \equal{\z}{0}}{
    	            \draw[\col] (-.5,1,0) -- (1,1,0);
    	            \draw[\col] (4,1,0) -- (5.5,1,0);
    	        }{
                    \draw[\col] (-.5,\y,\z) -- (5.5,\y,\z);
                }
    	    }
    	    \foreach \x in {0,1,2,3,4,5}{
                \draw[\col] (\x,-.5,\z) -- (\x,2.5,\z);
    	    }
	    }
	    \draw[wig] (1,1,0) -- (4,1,0);
	    \node[fill=white,circle,draw=black,inner sep=1.5pt] at (1,1,0) {};
	    \node[fill=white,circle,draw=black,inner sep=1.5pt] at (4,1,0) {};
	    \node[yshift=6pt,text=BrickRed] at (1.4,1,0) {${\sss \mc E_{1}}$};
	    \node[yshift=-5pt,xshift=-5pt,text=BrickRed] at (1,1,0) {${\sss \ldw}$};
	    \node[yshift=6pt,xshift=7pt,text=BrickRed] at (4,1,0) {${\sss \rdw}$};
	    \node[yshift=-6pt,text=BlueViolet] at (2.5,0,0) {${\sss n}$};
	\end{tikzpicture}
}

\newcommand{\condStringAlt}[0]{
	\begin{tikzpicture}[scale=1.1,baseline={([yshift=-.5ex]current bounding box.center)},z={(-0.4,0.4)}]
	    	    	 
        \draw[color=black!50] (-.5,0,1) -- (3.5,0,1);   
	    \foreach \x in {0,1,2,3}{
            \draw[color=black!50] (\x,0,0) -- (\x,0,1.5);
            \draw[color=black] (\x,-.5,0) -- (\x,.5,0);
            \draw[color=black!50] (\x,-.5,1) -- (\x,.5,1);
        }
	    \draw[wig] (-.5,0,0) -- (1,0,0);
	    \draw[wigb] (1,0,0) -- (2,0,0);
	    \draw[wig] (2,0,0) -- (3.5,0,0);
	    \node[yshift=6pt,text=BrickRed] at (0.4,0,0) {${\sss \mc E_{1}}$};
	    \node[yshift=-6pt,text=BlueViolet] at (1.5,0,0) {${\sss n}$};
	\end{tikzpicture}
}

\newcommand{\tinyOneCharge}[0]{
	\begin{tikzpicture}[scale=1.1,baseline={([yshift=-.5ex]current bounding box.center)},z={(-0.4,0.4)}]	    
	    \foreach \x in {0,1}{
	        \foreach \y in {0,1}{
	            \draw[color=black!50] (\x,\y,0) -- (\x,\y,1.5);
	        }
        }
	    \foreach \z/\col in {1/black!50,0/black}{
    	    \foreach \x in {0,1}{
    	        \ifthenelse{\equal{\z}{0}}{
    	            \draw[color=\col] (\x,-0.5,\z) -- (\x,0,\z);
    	            \draw[color=\col] (\x,1,\z) -- (\x,1.5,\z);
    	            \draw[color=\col] (-0.5,\x,\z) -- (0,\x,\z);
    	            \draw[color=\col] (1,\x,\z) -- (1.5,\x,\z);
    	        }{
                    \draw[color=\col] (\x,-0.5,\z) -- (\x,1.5,\z);
                    \draw[color=\col] (-.5,\x,\z) -- (1.5,\x,\z);
                }
    	    }
	    }
        \draw[wig] (0,0,0) -- (1,0,0);
        \draw[wig] (1,0,0) -- (1,1,0);
        \draw[wig] (1,1,0) -- (0,1,0);
        \draw[wig] (0,1,0) -- (0,0,0);
        \node at (.5,.5,0) {${\sss \Sigma_Z}$};
        \node[yshift=-8pt,text=BrickRed] at (0.5,0,0) {${\sss \mc E_{\omega}}$};
	\end{tikzpicture}
}

\newcommand{\OneCharge}[0]{
	\begin{tikzpicture}[scale=1.1,baseline={([yshift=-.5ex]current bounding box.center)},z={(-0.4,0.4)}]	    
	    \foreach \x in {0,1,2,3}{
	        \foreach \y in {0,1,2,3}{
	            \draw[color=black!50] (\x,\y,0) -- (\x,\y,1.5);
	        }
        }
	    \foreach \z/\col in {1/black!50,0/black}{
    	    \foreach \x in {0,1,2,3}{
    	        \ifthenelse{\(\equal{\x}{0} \OR \equal{\x}{3}\) \AND \equal{\z}{0}}{
    	            \draw[color=\col] (\x,-0.5,\z) -- (\x,0,\z);
    	            \draw[color=\col] (\x,3,\z) -- (\x,3.5,\z);
    	            \draw[color=\col] (-0.5,\x,\z) -- (0,\x,\z);
    	            \draw[color=\col] (3,\x,\z) -- (3.5,\x,\z);
    	        }{
                    \draw[color=\col] (\x,-0.5,\z) -- (\x,3.5,\z);
                    \draw[color=\col] (-.5,\x,\z) -- (3.5,\x,\z);
                }
    	    }
	    }
        \draw[wig] (0,0,0) -- (3,0,0);
        \draw[wig] (3,0,0) -- (3,3,0);
        \draw[wig] (3,3,0) -- (0,3,0);
        \draw[wig] (0,3,0) -- (0,0,0);
	    \foreach \x in {0,1,2}{
	        \foreach \y in {0,1,2}{
	            \node at (\x+.5,\y.5,0) {${\sss \Sigma_Z^k}$};
	        }
        }
        \draw[dot, shorten <= 1pt, shorten >= 4pt] (1,1.5,0) -- (1.5,2.5,0);
        \draw[dot, shorten <= 1pt, shorten >= 4pt] (1,.5,0) -- (1.5,1.5,0);
        \draw[dot, shorten <= 1pt, shorten >= 4pt] (1,.5,0) -- (1.5,2.5,0);
        \draw[dot, shorten <= 1pt, shorten >= 4pt] (2,1.5,0) -- (2.5,2.5,0);
        \draw[dot, shorten <= 1pt, shorten >= 4pt] (2,.5,0) -- (2.5,1.5,0);
        \draw[dot, shorten <= 1pt, shorten >= 4pt] (2,.5,0) -- (2.5,2.5,0);
        \node[fill=white,circle,inner sep=0pt] at (2,.5,0) {${\sss {\rm c}}$};
        \node[fill=white,circle,inner sep=0pt] at (2,1.5,0) {${\sss {\rm c}}$};
        \node[fill=white,circle,inner sep=0pt] at (1,.5,0) {${\sss {\rm c}}$};
        \node[fill=white,circle,inner sep=0pt] at (1,1.5,0) {${\sss {\rm c}}$};
        \node[yshift=-8pt,text=BrickRed] at (0.5,0,0) {${\sss \mc E_{\omega^k}}$};
	\end{tikzpicture}
}

\newcommand{\OneChargeMor}[0]{
	\begin{tikzpicture}[scale=1.1,baseline={([yshift=-.5ex]current bounding box.center)},z={(-0.4,0.4)}]	 
		    
	    \foreach \x in {0,1,2,3,4}{
	        \foreach \y in {1,2,3,4}{
	            \draw[color=black!50] (\x,\y,0) -- (\x,\y,1.5);
	        }
        }
        \foreach \x in {0,1,2,3,4}{
            \draw[color=black!50] (\x,0.5,1) -- (\x,4.5,1);
        }
        \foreach \x in {1,2,3,4}{
            \draw[color=black!50] (-.5,\x,1) -- (4.5,\x,1); 
        }
        \foreach \x in {1,2,3}{
            \draw[color=black] (\x,.5,0) -- (\x,4.5,0);
        }
        \foreach \y in {2,3}{
            \draw[color=black] (-.5,\y,0) -- (4.5,\y,0);
        }
        \foreach \x in {0,4}{
            \draw[color=black] (\x,.5,0) -- (\x,1,0);
            \draw[color=black] (\x,4,0) -- (\x,4.5,0);
        }
        \foreach \y in {1,4}{
            \draw[color=black] (-.5,\y,0) -- (0,\y,0);
            \draw[wigb] (1,\y,0) -- (2,\y,0);
            \draw[color=black] (4,\y,0) -- (4.5,\y,0);
        }
        
        \draw[wig] (0,1,0) -- (1,1,0);
        \draw[wig] (2,1,0) -- (4,1,0);
        \draw[wig] (4,1,0) -- (4,4,0);
        \draw[wig] (4,4,0) -- (2,4,0);
        \draw[wig] (1,4,0) -- (0,4,0);
        \draw[wig] (0,4,0) -- (0,1,0);
        
	    \foreach \x in {0,1,2,3}{
	        \foreach \y in {1,2,3}{
	            \ifthenelse{\x < 2}{
	                \node at (\x+.5,\y.5,0) {${\sss \Sigma_Z}$};
                }{
                    \node at (\x+.5,\y.5,0) {${\sss \Sigma_Z^\dagger}$};
                }
	        }
        }
        \foreach \x in {1,2,3}{
            \draw[dot, shorten <= 1pt, shorten >= 4pt] (\x,1.5,0) -- (\x+.5,2.5,0);
            \draw[dot, shorten <= 1pt, shorten >= 4pt] (\x,2.5,0) -- (\x+.5,3.5,0);
            \draw[dot, shorten <= 1pt, shorten >= 4pt] (\x,1.5,0) -- (\x+.5,3.5,0);
            \foreach \y in {1,2}{
                \node[fill=white,circle,inner sep=0pt] at (\x,\y+.5,0) {${\sss {\rm c}}$};
            }
        }
        \node[yshift=7pt,text=BrickRed] at (0.5,4,0) {${\sss \mc E_{\omega}}$};
        \node[yshift=-8pt,text=BrickRed] at (0.5,1,0) {${\sss \mc E_{\omega}}$};
        \node[yshift=-8pt,text=BrickRed] at (2.5,1,0) {${\sss \mc E_{\bar \omega}}$};
        \node[yshift=7pt,text=BrickRed] at (2.5,4,0) {${\sss \mc E_{\bar \omega}}$};
        \node[yshift=-8pt,text=BlueViolet] at (1.5,1,0) {${\sss n_{\omega}}$};
        \node[yshift=7pt,text=BlueViolet] at (1.5,4,0) {${\sss n_{\omega}}$};
	\end{tikzpicture}
}

\newcommand{\condStringMono}[1]{
	\begin{tikzpicture}[scale=1.1,baseline={([yshift=-.5ex]current bounding box.center)},z={(-0.4,0.4)}]
	    	    	    
	    \foreach \x in {0,1,2}{
	        \foreach \y in {0,1,2}{
	            \draw[color=black!50] (\x,\y,0) -- (\x,\y,1.5);
	        }
        }
	    \foreach \z/\col in {1/black!50,0/black}{
    	    \foreach \y in {0,1,2}{
    	        \ifthenelse{\equal{\y}{1} \AND \equal{\z}{0}}{
    	        }{
                    \draw[\col] (-.5,\y,\z) -- (2.5,\y,\z);
                }
    	    }
    	    \foreach \x in {0,1,2}{
                \draw[\col] (\x,-.5,\z) -- (\x,2.5,\z);
    	    }
	    }
	    \ifthenelse{#1=1}{
	        \node[fill=white,rectangle,inner sep=1pt] at (1,.55,0) {${\sss \sigma_X}$};
	        \node[fill=white,rectangle,inner sep=1pt] at (1,1.55,0) {${\sss \sigma_X}$};
	        \node[fill=white,rectangle,inner ysep=1pt,inner xsep=-1pt,text=black!50] at (1,1,.5) {${\sss \sigma_X}$};
	        \node at (1.5,1.5,0) {${\sss \Gamma}$};
    	    \fill[BlueViolet, fill opacity=0.15] (0,1,0) -- (2,1,0) -- (2,1,1) -- (0,1,1) -- (0,1,0);
    	    \fill[BlueViolet, fill opacity=0.15] (0,0,0) rectangle (2,2,0);
    	    \node[yshift=-6pt,text=BlueViolet] at (.5,0,0) {${\sss n}$};
    	    \node[yshift=-6pt,text=BlueViolet] at (1.5,0,0) {${\sss n}$};
	    }{}
	    \draw[wig] (-.5,1,0) -- (2.5,1,0);
	    \node[yshift=-6pt,text=BrickRed] at (-.2,1,0) {${\sss \mc E_{1}}$};
	    \node[yshift=-6pt,text=BrickRed] at (2.2,1,0) {${\sss \mc E_{1}}$};
	\end{tikzpicture}
}

\newcommand{\condStringComp}[1]{
	\begin{tikzpicture}[scale=1.1,baseline={([yshift=-.5ex]current bounding box.center)},z={(-0.4,0.4)}]
	    	    	    
	    \foreach \x in {0,1}{
	        \foreach \y in {0,1,2}{
	            \draw[color=black!50] (\x,\y,0) -- (\x,\y,1.5);
	        }
        }
	    \foreach \z/\col in {1/black!50,0/black}{
    	    \foreach \y in {0,1,2}{
    	        \ifthenelse{\equal{\y}{1} \AND \equal{\z}{0}}{
    	        }{
                    \draw[\col] (-.5,\y,\z) -- (1.5,\y,\z);
                }
    	    }
    	    \foreach \x in {0,1}{
                \draw[\col] (\x,-.5,\z) -- (\x,2.5,\z);
    	    }
	    }
	    \ifthenelse{#1=1}{
	        \draw (0,1,0) -- (1,1,0);
    	    \draw[wig] (-.5,1,0) -- (0,1,0);
    	    \draw[wig] (1,1,0) -- (1.5,1,0);
	        \node[fill=white,circle,draw=black,inner sep=1.5pt] at (0,1,0) {};
	        \node[fill=white,circle,draw=black,inner sep=1.5pt] at (1,1,0) {};
	        \node[yshift=-5pt,xshift=-6pt,text=BrickRed] at (0,1,0) {${\sss \rdw}$};
	        \node[yshift=6pt,xshift=7pt,text=BrickRed] at (1,1,0) {${\sss \ldw}$};
	    }{}
	    \ifthenelse{#1=2}{
	        \draw (0,1,0) -- (1,1,0);	    
    	    \draw[wig] (-.5,1,0) -- (0,1,0);
    	    \draw[wig] (1,1,0) -- (1.5,1,0);
	        \node[fill=white,circle,draw=black,inner sep=1.5pt] at (0,1,0) {};
	        \node[fill=white,circle,draw=black,inner sep=1.5pt] at (1,1,0) {};
	        \node[yshift=-5pt,xshift=-6pt,text=BrickRed] at (0,1,0) {${\sss \rdw}$};
	        \node[yshift=6pt,xshift=7pt,text=BrickRed] at (1,1,0) {${\sss \ldw}$};
            \node[text=black,inner sep=1pt,fill=white, rectangle] at (.5,1,0) {${\sss \uparrow}$};
	    }{}
	    \ifthenelse{#1=3}{
	        \draw (0,1,0) -- (1,1,0);	    
    	    \draw[wig] (-.5,1,0) -- (0,1,0);
    	    \draw[wig] (1,1,0) -- (1.5,1,0);
	        \node[fill=white,circle,draw=black,inner sep=1.5pt] at (0,1,0) {};
	        \node[fill=white,circle,draw=black,inner sep=1.5pt] at (1,1,0) {};
	        \node[yshift=-5pt,xshift=-6pt,text=BrickRed] at (0,1,0) {${\sss \rdw}$};
	        \node[yshift=6pt,xshift=7pt,text=BrickRed] at (1,1,0) {${\sss \ldw}$};
            \node[text=black,inner sep=1pt,fill=white, rectangle] at (.5,1,0) {${\sss \downarrow}$};
	    }{}
	    \ifthenelse{#1=4}{
            \draw[wig] (-.5,1,0) -- (1.5,1,0);
    	    \node[yshift=-6pt,text=BrickRed] at (-.2,1,0) {${\sss \mc E_{1}}$};
	    }{}
	    \ifthenelse{#1=5}{
    	    \node[yshift=-6pt,text=BrickRed] at (-.2,1,0) {${\sss \mc E_{1}}$};
            \fill[BlueViolet, fill opacity=0.15] (0,1,0) -- (1,1,0) -- (1,1,1) -- (0,1,1) -- (0,1,0);
            \fill[BlueViolet, fill opacity=0.15] (0,0,0) rectangle (1,2,0);
            \draw[wig] (-.5,1,0) -- (1.5,1,0);
            \node[yshift=-6pt,text=BlueViolet] at (.5,0,0) {${\sss n}$};
	    }{}
	\end{tikzpicture}
}

\newcommand{\OneChargeMorMono}[1]{
	\begin{tikzpicture}[scale=1.1,baseline={([yshift=-.5ex]current bounding box.center)},z={(-0.4,0.4)}]	 
		    
	    \foreach \x in {0,1,2,3,4,5}{
	        \foreach \y in {1,2,3,4}{
	            \draw[color=black!50] (\x,\y,0) -- (\x,\y,1.5);
	        }
        }
        \foreach \x in {0,1,2,3,4,5}{
            \draw[color=black!50] (\x,0.5,1) -- (\x,4.5,1);
        }
        \foreach \y in {1,2,3,4}{
            \draw[color=black!50] (-.5,\y,1) -- (5.5,\y,1); 
        }
        \foreach \x in {1,2,3,4}{
            \draw[color=black] (\x,.5,0) -- (\x,4.5,0);
        }
        \foreach \y in {2,3}{
            \draw[color=black] (-.5,\y,0) -- (5.5,\y,0);
        }
        \foreach \x in {0,5}{
            \draw[color=black] (\x,.5,0) -- (\x,1,0);
            \draw[color=black] (\x,4,0) -- (\x,4.5,0);
        }
        \foreach \y in {1,4}{
            \draw[color=black] (-.5,\y,0) -- (0,\y,0);
            \draw[wigb] (1,\y,0) -- (3,\y,0);
            \draw[color=black] (5,\y,0) -- (5.5,\y,0);
        }
        
        \draw[wig] (0,1,0) -- (1,1,0);
        \draw[wig] (3,1,0) -- (5,1,0);
        \draw[wig] (5,1,0) -- (5,4,0);
        \draw[wig] (5,4,0) -- (3,4,0);
        \draw[wig] (1,4,0) -- (0,4,0);
        \draw[wig] (0,4,0) -- (0,1,0);
        
	    \foreach \x in {0,1,2,3,4}{
	        \foreach \y in {1,2,3}{
	            \ifthenelse{\x < 2 \OR \x > 2}{
	                \node at (\x+.5,\y.5,0) {${\sss \Sigma_Z}$};
                }{
                    \node at (\x+.5,\y.5,0) {${\sss \Sigma_Z^\dagger}$};
                }
	        }
        }
        \foreach \x in {1,2,3,4}{
            \draw[dot, shorten <= 1pt, shorten >= 4pt] (\x,1.5,0) -- (\x+.5,2.5,0);
            \draw[dot, shorten <= 1pt, shorten >= 4pt] (\x,2.5,0) -- (\x+.5,3.5,0);
            \draw[dot, shorten <= 1pt, shorten >= 4pt] (\x,1.5,0) -- (\x+.5,3.5,0);
            \foreach \y in {1,2}{
                \node[fill=white,circle,inner sep=0pt] at (\x,\y+.5,0) {${\sss {\rm c}}$};
            }
        }
        \node[yshift=7pt,text=BrickRed] at (0.5,4,0) {${\sss \mc E_{\omega}}$};
        \node[yshift=-8pt,text=BrickRed] at (0.5,1,0) {${\sss \mc E_{\omega}}$};
        \node[yshift=7pt,text=BrickRed] at (4.5,4,0) {${\sss \mc E_{\omega}}$};
        \node[yshift=-8pt,text=BrickRed] at (4.5,1,0) {${\sss \mc E_{\omega}}$};
        \node[yshift=-8pt,text=BlueViolet] at (2.5,1,0) {${\sss n_{\bar \omega}}$};
        \node[yshift=7pt,text=BlueViolet] at (2.5,4,0) {${\sss n_{\bar \omega}}$};
        \node[yshift=-8pt,text=BlueViolet] at (1.5,1,0) {${\sss n_{\omega}}$};
        \node[yshift=7pt,text=BlueViolet] at (1.5,4,0) {${\sss n_{\omega}}$};
	\end{tikzpicture}
}

\newcommand{\OneChargeMonoidal}[1]{
	\begin{tikzpicture}[scale=1.1,baseline={([yshift=-.5ex]current bounding box.center)},z={(-0.4,0.4)}]
        
        \foreach \x in {0,1,2}{
            \foreach \y in {0,1,2}{
                \draw[black!50] (\x,\y,1) -- (\x,\y,1.5);
            }
        }
        \draw[black!50] (1,1,0) -- (1,1,1.5);
        \foreach \z/\col in {1/black!50,0/black}{
            \foreach \x in {0,1,2}{
                \ifthenelse{\equal{\x}{0} \OR \equal{\x}{2}}{
                    \draw[\col] (-.5,\x,\z) -- (0,\x,\z);
                    \draw[\col] (2,\x,\z) -- (2.5,\x,\z);
                    \draw[\col] (\x,-.5,\z) -- (\x,0,\z);
                    \draw[\col] (\x,2,\z) -- (\x,2.5,\z);
                }{
                    \draw[\col] (-.5,\x,\z) -- (2.5,\x,\z);
                    \draw[\col] (\x,-.5,\z) -- (\x,2.5,\z);
                }
            }
        }
                    
	    \ifthenelse{#1=1}{
	        \foreach \x/\y in {0/0,1/0,2/0,0/1,2/1,0/2,1/2,2/2}{
	            \draw[color=black!50] (\x,\y,0) -- (\x,\y,1);
	        }
	        \node[yshift=6pt,text=BrickRed] at (0.25,2,0) {${\sss \mc E_{\omega}}$};
	        \node[yshift=6pt,text=BrickRed!60] at (0.25,2,1) {${\sss \mc E_{\omega}}$};
        }{}
	    \ifthenelse{#1=2}{
	        \foreach \x/\y in {0/0,1/0,2/0,0/1,2/1,0/2,1/2,2/2}{
	            \draw[color=black!50] (\x,\y,0) -- (\x,\y,1);
	            \node[text=black!50,inner sep=0pt,fill=white, rectangle] at (\x,\y,.5) {${\sss \uparrow}$};
	        }
	        \node[yshift=6pt,text=BrickRed] at (0.25,2,0) {${\sss \mc E_{\omega}}$};
	        \node[yshift=6pt,text=BrickRed!60] at (0.25,2,1) {${\sss \mc E_{\omega}}$};
        }{}
        \ifthenelse{#1=3}{
	        \foreach \x/\y in {0/0,1/0,2/0,0/1,2/1,0/2,1/2,2/2}{
	            \draw[color=black!50] (\x,\y,0) -- (\x,\y,1);
	            \node[text=black!50,inner sep=0pt,fill=white, rectangle] at (\x,\y,.5) {${\sss \downarrow}$};
	        }
	        \node[yshift=6pt,text=BrickRed] at (0.25,2,0) {${\sss \mc E_{\omega}}$};
	        \node[yshift=6pt,text=BrickRed!60] at (0.25,2,1) {${\sss \mc E_{\omega}}$};
        }{}
        \ifthenelse{#1=4 \OR #1=5}{
	        \foreach \x/\y in {0/0,1/0,2/0,0/1,2/1,0/2,1/2,2/2}{
	            \draw[wig,opacity=0.5] (\x,\y,0) -- (\x,\y,1);
	        }
        }{}
        
        \foreach \z in {0,1}{
            \draw[wig, opacity = 1-0.5*\z] (0,0,\z) -- (2,0,\z);
            \draw[wig, opacity = 1-0.5*\z] (2,0,\z) -- (2,2,\z);
            \draw[wig, opacity = 1-0.5*\z] (2,2,\z) -- (0,2,\z);
            \draw[wig, opacity = 1-0.5*\z] (0,2,\z) -- (0,0,\z);
        }
        
        \begin{pgfonlayer}{back}
            \draw[dot, shorten <= 1pt, shorten >= 4pt,black!50] (1,.5,1) -- (1.5,1.5,1);
            \foreach \x/\y in {0/0,1/0,0/1,1/1}{
                \node[text=black!50] at (\x+.5,\y+.5,1) {${\sss \Sigma_Z}$};   
            }
        \end{pgfonlayer}
        
        \node[fill=white,circle,inner sep=0pt,text=black!50] at (1,.5,1) {${\sss {\rm c}}$};
        
        \ifthenelse{#1=1 \OR #1=2 \OR #1=3 \OR #1=4}{
            \draw[dot, shorten <= 1pt, shorten >= 4pt] (1,.5,0) -- (1.5,1.5,0);
            \node[fill=white,circle,inner sep=0pt] at (1,.5,0) {${\sss {\rm c}}$};
            \foreach \x/\y in {0/0,1/0,0/1,1/1}{
                \node[] at (\x+.5,\y+.5,0) {${\sss \Sigma_Z}$};   
            }
        }{
            \draw[dot, shorten <= 1pt, shorten >= 4pt] (1,.5,0) -- (1.5,1.5,0);
            \node[fill=white,circle,inner sep=0pt] at (1,.5,0) {${\sss {\rm c}}$};
            \foreach \x/\y in {0/0,1/0,0/1,1/1}{
                \node[] at (\x+.5,\y+.5,0) {${\sss \Sigma_Z^\dagger}$}; 
            }
        }
	\end{tikzpicture}
}

\newcommand{\OneChargeMonoidalCancel}[0]{
	\begin{tikzpicture}[scale=1.1,baseline={([yshift=-.5ex]current bounding box.center)},z={(-0.4,0.4)}]

        \foreach \x in {0,1,2,3,4,5}{
            \foreach \y in {0,1,2,3,4,5}{
                \ifthenelse {\(\(\equal{\y}{1} \OR \equal{\y}{4}\) \AND \(\x > 0  \AND \x < 5\)\) \OR \(\(\equal{\x}{1} \OR \equal {\x}{4}\) \AND \(\y > 1 \AND \y < 4\)\)}{
                }{
                    \draw[black!50] (\x,\y,0) -- (\x,\y,1.5);
                }
            }
        }
        \foreach \z/\col in {1/black!50,0/black}{
            \foreach \w in {0,2,3,5}{
                \draw[\col] (-.5,\w,\z) -- (5.5,\w,\z);
                \draw[\col] (\w,-.5,\z) -- (\w,5.5,\z);
            }
            \foreach \w in {1,4}{
                \draw[\col] (-.5,\w,\z) -- (1,\w,\z);
                \draw[\col] (4,\w,\z) -- (5.5,\w,\z);
                \draw[\col] (\w,-.5,\z) -- (\w,1,\z);
                \draw[\col] (\w,4,\z) -- (\w,5.5,\z);
            }
        }
        \foreach \z in {0,1}{
            \draw[wig, opacity = 1-0.5*\z] (1,1,\z) -- (4,1,\z);
            \draw[wig, opacity = 1-0.5*\z] (4,1,\z) -- (4,4,\z);
            \draw[wig, opacity = 1-0.5*\z] (4,4,\z) -- (1,4,\z);
            \draw[wig, opacity = 1-0.5*\z] (1,4,\z) -- (1,1,\z);
        }
        \foreach \x in {1,4}{
            \foreach \y in {1,2,3,4}{
                \draw[wig, opacity = 0.5] (\x,\y,0) -- (\x,\y,1);
            }
        }
        \foreach \x in {2,3}{
            \foreach \y in {1,4}{
                \draw[wig, opacity = 0.5] (\x,\y,0) -- (\x,\y,1);
            }
        }
        \foreach \x in {.5,4.5}{
            \foreach \y in {1,2,3,4}{
                \node[fill=white,rectangle,inner sep=1pt] at (\x,\y,0) {${\sss \sigma_X}$};
                \ifthenelse{\(\y > 1 \AND \y < 4\) \AND \equal{\x}{.5}}{
                    \node[fill=white,rectangle,inner sep=1pt] at ($(\x,\y,0)+(1,0,0)$) {${\sss \sigma_X}$};
                }{}
                \ifthenelse{\(\y > 1 \AND \y < 4\) \AND \equal{\x}{4.5}}{
                    \node[fill=white,rectangle,inner sep=1pt] at ($(\x,\y,0)+(-1,0,0)$) {${\sss \sigma_X}$};
                }{}
            }
        }
        \foreach \y in {.55,4.55}{
            \foreach \x in {1,2,3,4}{
                \node[fill=white,rectangle,inner sep=1pt] at (\x,\y,0) {${\sss \sigma_X}$};
                \ifthenelse{\(\x > 1 \AND \x < 4\) \AND \equal{\y}{.55}}{
                    \node[fill=white,rectangle,inner sep=1pt] at ($(\x,\y,0)+(0,1,0)$) {${\sss \sigma_X}$};
                }{}
                \ifthenelse{\(\x > 1 \AND \x < 4\) \AND \equal{\y}{4.55}}{
                    \node[fill=white,rectangle,inner sep=1pt] at ($(\x,\y,0)+(0,-1,0)$) {${\sss \sigma_X}$};
                }{}
            }
        }
        \foreach \w in {2,3}{
            \fill[BlueViolet, fill opacity=0.15] (0,\w,0) -- (2,\w,0) -- (2,\w,1) -- (0,\w,1) -- (0,\w,0);
            \fill[BlueViolet, fill opacity=0.15] (3,\w,0) -- (5,\w,0) -- (5,\w,1) -- (3,\w,1) -- (3,\w,0);
            \fill[BlueViolet, fill opacity=0.15] (\w,0,0) -- (\w,2,0) -- (\w,2,1) -- (\w,0,1) -- (\w,0,0);
            \fill[BlueViolet, fill opacity=0.15] (\w,3,0) -- (\w,5,0) -- (\w,5,1) -- (\w,3,1) -- (\w,3,0);
        }
        \foreach \w in {1,4}{
            \fill[BlueViolet, fill opacity=0.15] (0,\w,0) -- (1,\w,0) -- (1,\w,1) -- (0,\w,1) -- (0,\w,0);
            \fill[BlueViolet, fill opacity=0.15] (4,\w,0) -- (5,\w,0) -- (5,\w,1) -- (4,\w,1) -- (4,\w,0);
            \fill[BlueViolet, fill opacity=0.15] (\w,0,0) -- (\w,1,0) -- (\w,1,1) -- (\w,0,1) -- (\w,0,0);
            \fill[BlueViolet, fill opacity=0.15] (\w,4,0) -- (\w,5,0) -- (\w,5,1) -- (\w,4,1) -- (\w,4,0);
        }
        \foreach \x in {1,4}{
            \foreach \y in {1,2,3,4}{
                \node at (\x+.5,\y+.5,0) {${\sss \Gamma}$};
            }
        }
        \foreach \x in {2,3}{
            \foreach \y in {1,4}{
                \node at (\x+.5,\y+.5,0) {${\sss \Gamma}$};
            }
        }

        \node[yshift=6pt,text=BrickRed] at (1.25,4,0) {${\sss \mc E_{\omega}}$};
        \node[yshift=0pt,text=BrickRed!60] at (.75,3.25,1) {${\sss \mc E_{\omega}}$};
	\end{tikzpicture}
}

\begin{document} 

\title{A toy model for categorical charges}
\author{\textsc{Clement Delcamp}}
\email{clement.delcamp@ugent.be}
\affiliation{\vspace{1em}\it Max Planck Institute for the Physics of Complex Systems, N\"othnitzer Stra{\ss}e 38, 01187 Dresden, Germany}

\begin{abstract}
    \noindent 
    We consider a higher gauge topological model in three spatial dimensions whose input datum is a 2-group encoding the mixing of a 0-form $\mathbb Z_2$- and 1-form $\mathbb Z_3$-symmetry. We study the excitation content of the theory on the symmetry-preserving boundary. We show that boundary operators are organised into the fusion 2-category of 2-representations of the 2-group. These can be interpreted as categorical charges for an effective boundary model that inherits a global 2-group symmetry from the bulk topological order. Interestingly, we find that certain simple 2-representations are physically interpreted as composites of intrinsic excitations and condensation defects. 
    
    \bigskip
\end{abstract}

\maketitle

\section{Introduction}

\noindent
In modern parlance, ordinary (global) \emph{symmetry} operations leaving a physical system invariant, are referred to as \emph{0-form} symmetries. These are generated by operators supported on one-codimensional submanifolds of spacetime and labelled by elements of a \emph{group} so that fusion of symmetry operators obeys the multiplication rule of the group. Crucially, symmetry operators are \emph{topological} so that correlation functions are insensitive to deformations of the underlying submanifold that preserve its topology, unless it passes through \emph{charged operators}, i.e. operators supported on zero-dimensional manifolds that transform non-trivially in a \emph{representation} of the group. A special role is then played by \emph{irreducible} representations as they decompose the action of the group and provide meaningful sets of quantum numbers.

In recent years, the concept of symmetry has been generalised in various ways. A first generalisation is obtained by relaxing the requirement that operators are supported on one-codimensional submanifolds. This results in the notion of $q$-form symmetry that admits symmetry operators for every $q$-codimensional submanifolds of spacetime \cite{Gaiotto:2014kfa}. Symmetry operators are still labelled by group elements, but for $q > 1$ the corresponding group must now be \emph{abelian} \cite{Eckman1961StructureMI}. Furthermore, the topological nature of the symmetry operators requires the charged operators to be $q$-dimensional. These higher-form global symmetries have received widespread attention in high-energy physics \cite{Kapustin:2014gua,Sharpe:2015mja,Gaiotto:2017yup,Gaiotto:2017tne,Bhardwaj:2017xup,Tachikawa:2017gyf,Hofman:2018lfz,Hsin:2018vcg,Delcamp:2019fdp,Sogabe:2019gif,Seiberg:2020bhn,Seiberg:2020cxy,Iqbal:2021rkn} and condensed matter theory \cite{Nussinov:2006iva,Nussinov:2009zz,Wen:2018zux,Lake:2018dqm,Zhao:2020vdn,Iqbal:2020msy,Rayhaun:2021ocs,Bi:2019ers,McGreevy:2022oyu}. A further generalisation consists in relaxing the \emph{invertibility} condition of symmetry operators so that fusion rules are encoded into abstractly higher mathematical structures such as \emph{fusion categories} \cite{etingofFusion}. These so-called \emph{categorical symmetries} have also been under scrutiny as of late \cite{PhysRevLett.98.160409,PhysRevLett.101.050401,PhysRevLett.103.070401,Kapustin:2009av,Kapustin:2010if,ardonneAnyonChain,PhysRevB.87.235120,Buican:2017rxc,Thorngren:2019iar,Huang:2021ytb,Thorngren:2021yso,Kaidi:2021xfk,Choi:2021kmx,tantivasadakarn2021longrange}. 
One concrete application of these generalised notions of symmetry is the broadening of Landau's theory of \emph{symmetry breaking} so as to include phases of matter previously believed not to be accommodated by it, e.g. \emph{topological phases} \cite{doi:10.1063/1.1665530,PhysRevLett.50.1395,Wen:1989iv,PhysRevB.82.155138}. 

Interestingly, higher-form symmetries of various degrees can be combined in non-trivial ways so that they do not factorise. A prototypical example is the mixing of 0- and 1-form symmetries, whereby the 0-form symmetry acts on the 1-form charged operators. The associated groups combine into a categorification of the notion of group referred to as a \emph{2-group} \cite{Sinh}. Much progress has been made towards understanding the resulting 2-group global symmetries and the corresponding anomalies \cite{Kapustin:2013uxa,Delcamp:2018wlb,Cordova:2018cvg,Benini:2018reh,Hsin:2019fhf,Iqbal:2020lrt}.
Nevertheless, the corresponding \emph{higher representation theory} in connection with the study of charged operators remains somewhat elusive.

There is indeed a generalised notion of representation that is suitable to 2-groups. These so-called \emph{2-representations} are defined as \emph{2-functors} from the 2-group---thought as a one-object \emph{2-groupoid}---to the 2-category of \emph{2-vector spaces} \cite{Crane:2003gk,Barrett:2004zb,Elgueta:2004}, categorifying the definition of a group (1-)representation as a functor from the group---thought as a one-object (1-)groupoid---to the category of vector spaces. Since every group is in particular a 2-group, it makes sense to compute its 2-representations. This special kind of 2-representations has been studied in detail in the past \cite{2002math......2130O,2006math......2510G,Bartlett}, and was recently shown to naturally arise in the context of the electromagnetic duality of higher-dimensional symmetric quantum models \cite{PhysRevResearch.2.033417,PhysRevResearch.2.043086,CDmod2}. Perhaps surprisingly, the 2-category of 2-representations of a group contains more information than its lower-categorical counterpart, encoding in particular so-called \emph{condensation defects} that arise for instance when gauging higher-form symmetries along submanifolds \cite{Kong:2014qka,PhysRevB.96.045136,gaiotto2019condensations,Kong:2019brm,Kong:2020wmn,Johnson-Freyd:2020twl,Roumpedakis:2022aik}.

In this work, we consider an effective model with a specific global 2-group symmetry and elucidate in this context the physical interpretation of the corresponding 2-representations. Our approach exploits a correspondence between \emph{symmetry-preserving} gapped phases in (2+1)d and \emph{Neumann} boundary conditions of (3+1)d \emph{topological} models satisfying gauged versions of the symmetries \cite{Thorngren:2019iar}. A non-anomalous global 2-group symmetry can indeed be gauged by coupling it to 1- and 2-form connections that interact in a non-trivial way. The resulting theory is typically referred to as a \emph{higher gauge theory} \cite{2003math......7200B,2005math.....11710B,Baez:2010ya}. 
Practically, we study an exactly solvable model with a higher gauge theory interpretation that provides a lattice Hamiltonian realisation of \emph{Yetter's homotopy 2-type} topological quantum field theory \cite{Yetter:1993dh}. The input datum is a strict 2-group encoding the mixing of a 0-form $\mathbb Z_2$-symmetry and 1-form $\mathbb Z_3$-symmetry. This model yields bulk charge and flux composite \emph{intrinsic} excitations that are either point-like or loop-like. The Neumann gapped boundary condition is then obtained by condensing all the magnetic excitations, and the resulting boundary operators can be identified with charged operators for the effective symmetric model. We explain how these \emph{categorical charges} and their fusion  statistics are organised into the monoidal bicategory of 2-representations of the 2-group. Interestingly, certain simple 2-representations are interpreted as composites of intrinsic excitations and condensation defects.

\section{Categorical charges}

\noindent
\emph{We propose in this section a toy model for categorical charges. These charges arise as boundary topological excitations of a lattice Hamiltonian possessing a higher gauge theory interpretation.}

\subsection{Lattice Hamiltonian}

\noindent
Let us consider an oriented three-dimensional manifold $\mc M$ with a non-empty boundary $\partial \mc M$. For now we assume that the boundary has a single connected component. We further suppose that $\mc M$ can be endowed with a \emph{cubic} cellulation $\mc M_\boxempty$ whose vertices, edges, plaquettes and cubes are notated via $\msf v$, $\msf e$, $\msf p$ and $\msf c$, respectively:
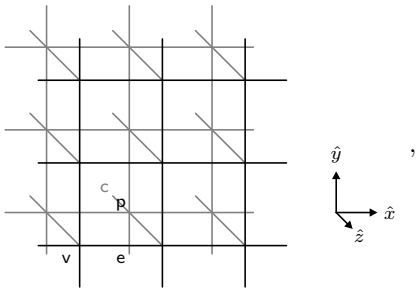
\begin{equation*}
    \lattice \, ,
\end{equation*}
where bold lines correspond to edges along the boundary $\partial \mc M$. We require parallel edges to be all oriented in the positive direction according to the frame of reference depicted above, and denote by ${\rm s}(\msf e)$ and ${\rm t}(\msf e)$ the source and target vertices of the directed edge $\msf e \subset \mc M_\boxempty$, respectively. This in turn induces an orientation for every plaquette $\msf p \subset \mc M_\boxempty$, and canonically assigns to it a \emph{basepoint} ${\rm bp}(\msf p)$ defined as the unique vertex $\msf v \subset \partial \msf p$ such that ${\rm s}(\msf e_1) = \msf v = {\rm s}(\msf e_2)$ for $\msf e_1, \msf e_2 \subset \partial \msf p$.

\emph{Qubit} and \emph{qutrit} degrees of freedom are assigned to edges and plaquettes of $\mc M_\boxempty$, respectively. We identify such an assignment with a choice of \emph{colouring} of the edges and plaquettes by $\mathbb Z_2$ and $\mathbb Z_3$ group variables, respectively, so that the microscopic Hilbert space is provided by $(\bigotimes_\msf e \mathbb C[\mathbb Z_2]) \otimes (\bigotimes_\msf p \mathbb C[\mathbb Z_3])$. We choose the presentations $\mathbb Z_2 = (\{+1,-1\},\times) \ni g$ and $\mathbb Z_3 = (\{0,1,2\},+) \ni h$, where the addition is modulo 3. Given a colouring of the cellulation we notate its restriction to an edge $\msf e$ or a plaquette $\msf p$ via $| g_\msf e \ra$ and $| h_\msf p \ra$, respectively, both regarded as elements of the microscopic Hilbert space. We identify by convention a qubit in the `up' state $| \uparrow \, \ra$ with $| +1 \ra$ or in the `down' state $| \downarrow \, \ra$ with $|-1\ra$. 

Qubit and qutrit degrees of freedom are governed by a lattice Hamiltonian obtained as a sum of mutually \emph{commuting} local operators split in four families. Let us begin by introducing some definitions. We require the following operators:
\begin{equation*}
    \sigma_X = 
    \begin{pmatrix}
        0 & 1
        \\
        1 & 0
    \end{pmatrix} 
    , \q
    \sigma_Z =
    \begin{pmatrix}
        1 & 0 
        \\
        0 & -1
    \end{pmatrix}
\end{equation*}
such that $\sigma_X \sigma_Z = - \sigma_Z \sigma_X$
and
\begin{equation*}
    \Sigma_X =
    \begin{pmatrix}
        0 & 0 & 1
        \\
        1 & 0 & 0
        \\
        0 & 1 & 0
    \end{pmatrix}
    \! , \q
    \Sigma_Z =
    \begin{pmatrix}
        1 & 0 & 0
        \\
        0 & \omega & 0
        \\
        0 & 0 & \bar \omega
    \end{pmatrix}
    \! , \q
    \Gamma =
    \begin{pmatrix}
        1 & 0 & 0
        \\
        0 & 0 & 1
        \\
        0 & 1 & 0
    \end{pmatrix} \! ,
\end{equation*}
with $\omega = \exp(\frac{2\pi i}{3})$ such that $\Sigma_Z^3 = \Sigma_X^3 = {\rm id}$, $\Sigma_Z \Sigma_X = \omega \Sigma_X \Sigma_Z$, $\Sigma_Z^\dagger \Sigma_X = \bar \omega \Sigma_X \Sigma_Z^\dagger$, $\Gamma \Sigma_Z \Gamma = \Sigma_Z^\dagger$ and $\Gamma \Sigma_X \Gamma = \Sigma_X^\dagger$. We also need \emph{controlled gates} of the form
\begin{align}
    \arraycolsep=1.4pt
    \begin{array}{ccccl}
        \cSX  & : & \mathbb C^2 \otimes \mathbb C^3 & \to & \hspace{1pt} \mathbb C^2 \otimes \mathbb C^3
        \\
        & : & | g,h \ra  & \mapsto & ({\rm id} \otimes \Sigma_X^{g})|g,h\ra
    \end{array} \, ,
\end{align}
where the qubit plays the role of the control. In practice we shall apply this gate between a qutrit assigned to a plaquette $\msf p$ and a qubit assigned to an edge $\msf e$ that may or may not be in the boundary of $\msf p$. We similarly define a controlled gate ${\rm c}\Sigma_Z$.  It immediately follows from the definitions that the following commutation relations are satisfied: 
\begin{align}
    \nn
    \cSX ({\rm id} \otimes \Gamma) &= ({\rm id} \otimes \Gamma)\cSX^\dagger \, ,
   \\
   \label{eq:commutRelationsCSX}
   \cSX (\sigma_X \otimes {\rm id}) &= (\sigma_X \otimes {\rm id}) \cSX^\dagger \, ,
   \\[2pt]
   \nn
   {\rm c}\Sigma_X (\sigma_X \otimes \Gamma) &= (\sigma_X \otimes \Gamma){\rm c}\Sigma_X \, .
\end{align}
Similar commutation relations hold for $\cSZ$. Moreover, we have $\cSZ \cSX = \omega \, \cSX \cSZ$, and generalisation thereof. We graphically represent such controlled gates by means of a dotted line connecting a control qubit identified by `c' and a target qutrit, e.g.

\begin{align*}
    \cGate{1} \; : \cGate{2} \; \mapsto \!\! \cGate{3} \, ,
    \\
    \cGate{4} \; : \cGate{2} \; \mapsto \!\! \cGate{5} \, ,
\end{align*}
for any $g \in \mathbb Z_2$ and $h \in \mathbb Z_3$, with $[g+h] \equiv (g+h)$ mod $3$. We shall also consider generalisations of these controlled gates with more than one control qubits, e.g. ${\rm cc}\Sigma_X : |g_1,g_2,h \ra \mapsto ({\rm id} \otimes {\rm id} \otimes \Sigma_X^{g_1g_2})|g_1,g_2,h \ra$, depicted as before via dotted lines connecting all the control qubits to the single qutrit.

We are now ready to write the lattice Hamiltonian, which we decompose as $\mathbb H = \mathbb H_{{\rm int}(\mc M)} + \mathbb H_{\partial \mc M}$, where $\mathbb H_{{\rm int}(\mc M)}$ only contains terms that act solely on qubit and qutrit degrees of freedom in the interior ${\rm int}(\mc M)$ of $\mc M$, and $\mathbb H_{\partial \mc M}$ only contains terms involving boundary degrees of freedom. Since we are ultimately interested in the boundary excitations, we shall only write down explicitly the component $\mathbb H_{\partial \mc M}$. Graphically, we have
\begin{align}   
    \nn
    \mathbb H_{\partial \mc M} = 
    &- \sum_{\msf p \bot \hat z} \, \hamOp{1}
    - \sum_{\msf p \bot \hat y} \, \hamOp{2}
    - \sum_{\msf p \bot \hat x} \, \hamOp{3}
    \\[-2pt] \nn
    &-\sum_{\msf v} \, \hamOp{4}
    \\ \nn
    &-\sum_{\msf c} \Bigg(\, \hamOp{9} + \hamOp{10} \, \Bigg)\Big( \! \prod_{\msf p \subset \partial \msf c} \! \delta(g_\msf p)\Big)
    \\ \nn
    &- \sum_{\msf e \parallel \hat y} \Bigg( \raisebox{6pt}{\hamOp{5}} 
    + \raisebox{6pt}{\hamOp{7}} \Bigg)
    \\
    &- \sum_{\msf e \parallel \hat x} \Bigg(\,  \hamOp{6} \;\, + \hamOp{8} \; \Bigg) \, ,
\end{align}
where $g_\msf p := \prod_{\msf e \subset \partial \msf p}g_\msf e$. We can immediately check that $\mathbb H_{\partial \mc M}$ is Hermitian. As mentioned earlier, we distinguish four families of local operators, which act on neighbourhoods of the plaquettes, vertices, cubes and edges of the cubic lattice $\mc M_\boxempty$, respectively. We interpret these operators as enforcing \emph{magnetic} and \emph{electric} stabiliser conditions for the qubit and qutrit degrees of freedom, respectively. 
Notice, however, that except for the plaquette operators, all operators involve both qubit and qutrit degrees of freedom.
We provide additional comments regarding the physical interpretations of these various
terms in sec.~\ref{sec:higherGauge}.

Let us confirm that all the operators entering the definition of $\mathbb H_{\partial \mc M}$ are mutually commuting. First of all, the plaquette operators readily commute with all operators. Vertex operators commute with any cube operator in virtue of $(\sigma_X \otimes \Gamma)\cSZ^g = \cSZ^g (\sigma_X \otimes \Gamma)$ and with the sum of any edge operator and its Hermitian conjugate either in virtue of $(\sigma_X \otimes \Gamma)\cSX^g = \cSX^g (\sigma_X \otimes \Gamma)$ or $(\sigma_X \otimes {\rm id})\cSX^g = \cSX^{-g} (\sigma_X \otimes {\rm id})$ together with $\Gamma \Sigma_X^{g}\Gamma = \Sigma_X^{-g}$. Finally, the commutation between cube and edge operators is guaranteed by $\Sigma_X^{g_1} \Sigma_Z^{g_2} = \omega^{g_1g_2} \Sigma_Z^{g_2} \Sigma_X^{g_1}$ as well as the fact that cube operators vanish whenever the plaquette stabiliser conditions are not satisfied. Let us emphasise that this last requirement is crucial whenever the commutation relation between an edge and a plaquette operator involves four controlled gates whose control qubits are associated with boundary edges of the same plaquette. All the operators commute and thus the model is exactly solvable. Note finally that when discarding the qutrit degrees of freedom, the model reduces to the usual (3+1)d toric code \cite{KITAEV20032,PhysRevB.72.035307}.

\subsection{Higher gauge theory interpretation \label{sec:higherGauge}}

\noindent
Let us now reveal the higher gauge theory interpretation of the Hamiltonian we consider. Loosely speaking, a higher gauge theory is a gauge theory for which the group is replaced by a \emph{categorification} of the notion of group referred to as a \emph{2-group}. We focus in this manuscript on \emph{strict} 2-groups defined as strict monoidal categories whose objects and morphisms are invertible \cite{Sinh,2003math......7200B}. Crucially strict 2-groups can be equivalently described in terms of \emph{crossed modules} \cite{bams/1183513543}. A crossed module $\mathbb G$ is a quadruple $(G,H,\act, \partial)$ consisting of two groups $G$ and $H$, a group action $\act: G \times H \to H$ by automorphisms, and a group homomorphism $\partial: G \to H$, which are subject to the so-called \emph{Peiffer} identities. Our attention is on the so-called \emph{dihedral} 2-group $\mathbb G(2,3) \equiv (\mathbb Z_2,\mathbb Z_3, \mathbb 1_{\mathbb Z_2}: \mathbb Z_3 \to \{+1\},\act)$, where $g \act h = h^{g}$ for any $g \in \mathbb Z_2$ and $h \in \mathbb Z_3$, so that $\mathbb Z_2 \ltimes \mathbb Z_3$ is isomorphic to the dihedral group $\mathbb D_6$ of order 6. The Peiffer identities being trivially satisfied for $\mathbb G(2,3)$, we need not writing them down explicitly. We can conveniently rewrite the controlled gates in terms of the action $\act$ as
\begin{equation}
\begin{split}
    \cSX^k &: |g,h \ra \mapsto | g, [h+(g \act k)] \ra
    \\
    \cSZ^k &: |g,h \ra \mapsto (\omega^h)^{g \act k}|g,h \ra \, ,
\end{split}
\end{equation}
for any $k \in \mathbb Z_3$.

Assigning group variables $g_\msf e \in \mathbb Z_2$ to every edge $\msf e \subset \mc M_\boxempty$ defines an (ordinary) 1-form $\mathbb Z_2$-connection, whereas assigning group variables $g_\msf p$ to every plaquette $\msf p \subset \mc M_\boxempty$ defines a 2-form $\mathbb Z_3$-connection. Together with the 1-flatness condition encoded into the plaquette operators, these assignments define a so-called \emph{2-connection} on the strict 2-group $\mathbb G(2,3)$. The cube operators then enforce a 2-flatness constraint on this 2-connection. Such a gauge field configuration is subject to two kinds of gauge transformations: On the one hand, we have ordinary (0-form) $\mathbb Z_2$-gauge transformations whose actions on the 2-connection are encoded into the vertex operators. On the other hand, there are 1-form $\mathbb Z_3$-gauge transformations acting on the 2-connection via the edge operators. Note that 0-form and 1-form gauge transformations typically do not commute, but commutation of the operators in the definition of the Hamiltonian is achieved by considering an edge operator together with its Hermitian conjugate. A general gauge transformation can then be defined as a 0-form transformation preceded by a 1-form transformation. Composition of such general gauge transformations is then provided by group multiplications in $\mathbb D_6$. The model we consider is an example of higher gauge models of topological phases considered in ref.~\cite{PhysRevB.95.155118,Bullivant:2019tbp}.

Let us remark that the higher gauge model with input datum $\mathbb G(2,3)$ admits by no means a unique choice of gapped boundary condition. The boundary Hamiltonian $\mathbb H_{\partial \mc M}$ we focus on in this manuscript encodes the so-called `smooth' (or Neumann) boundary condition. This boundary condition is obtained by condensing all the magnetic excitations, i.e. violations of 1- and 2-flatness constraints, so that only electric excitations, i.e. violations of 0- and 1-form gauge invariance, live at the boundary. The remainder of this manuscript is dedicated to the study of these electric excitations.

\subsection{Boundary excitations}

\noindent
Our model yields four types of bulk excitations obtained by violating stabiliser conditions associated with each family of Hamiltonian operators. In light of the higher gauge theory interpretation of the model, we shall refer to the corresponding four types of \emph{elementary} bulk excitations as \emph{1-flux}, \emph{0-charge}, \emph{2-flux} and \emph{1-charge}, respectively. We know from general considerations that 0-charges and 2-fluxes should be zero-dimensional objects, whereas 1-fluxes and 1-charges should be one-dimensional. 

We focus in this manuscript on topological boundary excitations for the boundary condition encoded into our choice of Hamiltonian $\mathbb H_{\partial \mc M}$. As evoked above, this is the boundary condition that condenses 1- and 2-fluxes. To precise this statement, we first need to clarify the notion of `condensation' within our context. For two-dimensional topological models, condensing a topological excitation amounts to making it \emph{local}---in the sense that it can be created by a local operator. However, this convention becomes ambiguous in higher dimensions. Consider for instance the (3+1)d toric code that hosts point-like (0-)charges and loop-like (1-)fluxes created at the boundary of string- and membrane-like topological operators, respectively. The `rough' (or Dirichlet) boundary condition condenses electric excitations since a charge brought to the boundary can be annihilated by a local boundary operator. What about the `smooth' (or Neumann) boundary condition? When bringing a loop-like flux to the boundary, it cannot be annihilated by a local operator but by a \emph{genuine line-like operator}, i.e. an operator that does \emph{not} depend on the topology of a choice of bounded surface. Equivalently, a loop-like flux excitation can be created at the boundary with a genuine line-like operator but it cannot be detected as we cannot link it to a topological charge, and thus it is identified with the vacuum. Going back to the higher gauge model of interest, we could show that the boundary condition under consideration condenses the magnetic excitations as 1-fluxes can be annihilated at the boundary by genuine line-like operators and 2-fluxes by local operators. This also means that 1-fluxes are defined up to genuine line-like operators. We shall exploit this property in the following but for now let us explicitly identify the boundary (electric) operators. 

\bigskip \noindent
Firstly, we denote by $\mathbb 1$ the trivial (string-like) 1-charge and by $1_\mathbb 1$ the trivial (point-like) 0-charge. As suggested by the notation, we shall think of a trivial 0-charge $1_\mathbb 1$ as being attached to a trivial 1-charge $\mathbb 1$, or rather as being a domain wall $\mathbb 1 \to \mathbb 1$ between two trivial 1-charges. The first non-trivial excitation is the 0-charge $e$. Akin to the (3+1)d toric code, pairs of 0-charges $e$ can be created at the endpoints of a string of $\sigma_Z$ operators, e.g.
\begin{equation}
    \ZeroCharge \, .
\end{equation}
As before, we think of a 0-charge $e$ as being a domain wall $e:\mathbb 1 \to \mathbb 1$ between two trivial 1-charges $\mathbb 1$. From these 0-charges descends another type of object, namely one-dimensional defects that condense 0-charges $e$. These \emph{condensation defects} do not constitute excitations of the Hamiltonian per se, but rather deformations of the model itself. Physically, such a defect is interpreted as a one-dimensional gas of 0-charges $e$ that spontaneously breaks the effective $\mathbb Z_2$-symmetry \cite{Kong:2014qka,PhysRevB.96.045136,Kong:2020wmn,Johnson-Freyd:2020twl}. Microscopically, it can be realised as follows: Given a (1-)path $\ell$ along the edges of the cubic lattice, we discard the vertex operators associated with the vertices $\msf v \subset \ell$ along the path, and add the energy term
\begin{equation}
    -\sum_{\msf e \subset \ell} \hamOp{0} \, ,
\end{equation}
which effectively projects the qubits along the path $\ell$ onto the spin-$\uparrow$ subspace. With respect to the original Hamiltonian, this defect amounts to creating at every edge along the path an equal-weight superposition of the vacuum and the excited state. We can then readily show that as we bring a $0$-charge $e$ in contact with this defect, it cannot be distinguished from the vacuum anymore. Henceforth, we denote this defect by $\mc E_{1}$ and depict it as a wiggly red line $\wigLine{1}$. Importantly, this string-like defect does not have to be closed, i.e. there exist domain walls between the trivial 1-charge $\mathbb 1$ and the condensation defect $\mc E_1$. We notate via $\ldw : \mathbb 1 \to \mc E_1$ and $\rdw: \mc E_1 \to \mathbb 1$ the (unique) domain walls associated with each endpoint of the string. Furthermore, we distinguish two types of domain walls between two condensation defects $\mc E_1$. Indeed, in addition to the trivial domain wall $1_{\mc E_1}: \mc E_1 \to \mc E_1$, there is a domain wall $n : \mc E_1 \to \mc E_1$ that corresponds to violating the 1-flatness stabiliser conditions associated with the three plaquettes meeting at an edge $\msf e \subset \ell$ along the path. 
We graphically summarise these various objects as
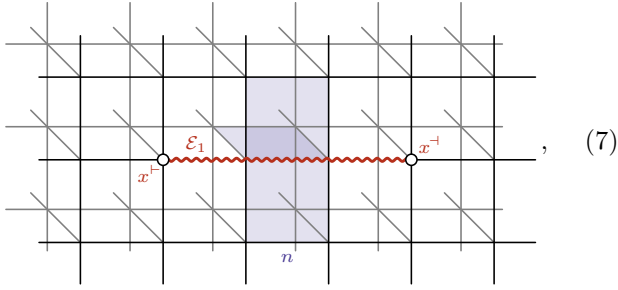
\begin{equation}
    \label{eq:condString}
    \condString \, ,
\end{equation}
where the blue-shaded plaquettes are those for which the 1-flatness/plaquette stabiliser conditions are violated. Note however that interpreting this excitation strictly as a 1-flux requires some care as the cube operators are defined to vanish whenever the 1-flatness is not everywhere satisfied around the cube (see \cite{Huxford:2022qtg} for a discussion on the interplay between 1- and 2-fluxes). The domain wall $n$ is distinct from the trivial one $1_{\mc E_1}$ since we cannot annihilate the resulting magnetic excitation with a local or genuine line-like operator without creating additional excitations or altering the condensation defect itself. Indeed, the only local operation that could annihilate this magnetic excitation would amount to modifying the defect by projecting the qubit at the intersection of the blue-shaded plaquettes onto the spin-$\downarrow$ subspace. This provides an alternative interpretation for the domain wall $n$ as 
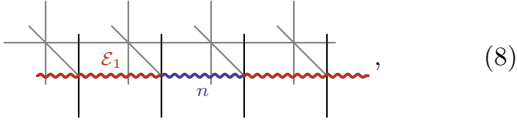
\begin{equation}
    \label{eq:altCondDefect}
    \condStringAlt \, ,
\end{equation}
where $\wigLine{2}$ indicates an energy term in the Hamiltonian effectively projecting the corresponding qubit onto the spin-$\downarrow$ subspace. We exploit both interpretations in the following.

\bigskip \noindent
The defects identified so far coincide with those found on the smooth boundary of the (3+1)d toric code by Kong et al. in \cite{Kong:2020wmn}. Let us now study another type of electric excitations, which is obtained by violating the 1-form gauge invariance. Consider inserting a $\Sigma_Z$ operator on a given plaquette $\msf p$. Such an operator insertion certainly breaks the edge stabiliser conditions at every $\msf e \subset \partial \msf p$. However, these violations do not correspond to elementary excitations, as witnessed by the fact that the commutator of $\Sigma_Z$ with two of the edge operators along $\partial \msf p$ depends on the $\mathbb Z_2$-colouring of two edges $\msf e_1, \msf e_2 \subset \partial \msf p$, as well as the fact that it does not commute with the vertex operator associated with ${\rm bp}(\msf p)$. This is remedied by projecting these boundary qubits onto a specific state. More precisely, we construct the 1-charge $\mc E_\omega$ by combining a $\Sigma_Z$ operator with a condensation defect running along the boundary of $\msf p$ so as to project the corresponding qubits onto the spin-$\uparrow$ subspace, i.e
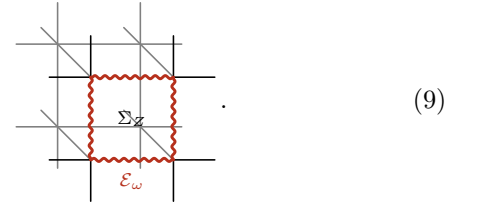
\begin{equation}
    \tinyOneCharge \, .
\end{equation}
Replacing $\Sigma_Z$ by $\Sigma^\dagger_Z$ yields another 1-charge denoted by $\mc E_{\bar \omega}$.
How about defining larger excitations? Given a 2-path $\mc P$, acting with $(\Sigma_Z^k)_\msf p$, with $k \in \mathbb Z_3$, at every $\msf p \subset \mc P$ would not yield the expected excitations. Indeed, given two adjoining plaquettes $\msf p_1$ and $\msf p_2$, the edge stabiliser operator acting at $\msf e = \msf p_1 \cap \msf p_2$ may or may not commute with $(\Sigma_Z^k)_{\msf p_1} \otimes (\Sigma_Z^k)_{\msf p_2}$ depending on the $\mathbb Z_2$-colouring of the edges bounding these plaquettes. More precisely, we are looking for a topological membrane operator that only creates a loop-like excitation at its boundary. We must therefore adapt this naive attempt in order to account for the dependence of the edge stabiliser operators on the qubits living on the 2-path. Assigning a basepoint ${\rm bp}(\mc P)$ to the 2-path $\mc P$ and denoting by $\ell_\mc P^\msf p : {\rm bp}(\mc P) \to {\rm bp}(\msf p)$ a choice of 1-path from ${\rm bp}(\mc P)$ to the basepoint of any plaquette $\msf p \subset \mc P$, we consider instead the operator\footnote{Analogous operators for more general higher gauge models were constructed in \cite{Huxford:2022wih,Huxford:2022mgr}.}
\begin{align}
    \label{eq:oneChargeOp}
    \prod_{\msf p \subset \mc P} \prod_{\msf e \subset \ell^\msf p_\mc P} (\cSZ^k)_{\msf e,\msf p} 
    =
    \prod_{\msf p \subset \mc P}(\Sigma_Z^k)_{\msf p}^{\prod_{\msf e \subset \ell^\msf p_\mc P}(\sigma_Z)_\msf e} 
    \, ,
\end{align}
with $k \in \mathbb Z_3$.
Concretely, this operator acts on a state associated with a plaquette $\msf p $ as
\begin{align*}
    |g_{\msf e_1}, \ldots, g_{\msf e_{|\ell^\msf p_\mc P|}}, h_\msf p\ra 
    \mapsto 
    & \; (\Sigma_Z^k)_\msf p^{\prod_{\msf e \subset \ell^\msf p_\mc P} g_{\msf e}}
    |g_{\msf e_1}, \ldots, g_{\msf e_{|\ell^\msf p_\mc P|}}, h_\msf p\ra 
    \\
    & = (\Sigma_Z^{g_{\ell^\msf p_\mc P} \! \act k})_\msf p
    |g_{\msf e_1}, \ldots, g_{\msf e_{|\ell^\msf p_\mc P|}}, h_\msf p\ra 
    \\
    & = (\omega^{h_\msf p})^{(g_{\ell^\msf p_\mc P} \act k)}
    |g_{\msf e_1}, \ldots, g_{\msf e_{|\ell^\msf p_\mc P|}}, h_\msf p\ra \, ,
\end{align*}
where $\msf e_1, \ldots, \msf e_{|\ell^\msf p_\mc P|}$ are the constitutive edges of the path $\ell^\msf p_\mc P$ and $g_{\ell^\msf p_\mc P} = \prod_{\msf e \subset \ell^\msf p_\mc P}g_{\msf e}$.
As long as the 1-flatness stabiliser conditions are everywhere satisfied, this operator is insensitive to the choices of 1-paths ${\rm bp}(\mc P) \to {\rm bp}(\msf p)$ for every $\msf p \subset \mc P$.
In practice, we shall focus on rectangular membrane-like operators for convenience, in which case we can always choose without loss of generality the basepoint ${\rm bp}(\mc P)$ of the 2-path $\mc P$ to be the basepoint of the bottom left plaquette. Combined with condensation defects running along the boundary $\partial \mc P$ of the 2-path, the operator $\eqref{eq:oneChargeOp}$ creates a loop-like 1-charge $\mc E_{\omega^k}$. For instance, we have
\begin{equation}
    \label{eq:exampleOneCharge}
    \OneCharge \, ,
\end{equation}
where we exploited the fact that qubits along the condensation defect are projected onto the spin-$\uparrow$ subspace to simplify the operators. Let us confirm that this defect excitation yields a well-defined 1-charge. In particular, we need to check that all the stabiliser conditions in the interior of $\mc P$ are fulfilled. Typically, we cannot readily apply the commutation relations between the $\cSX$ and $\cSZ$ gates as they are defined with respect to different control qubits. This is remedied by exploiting the fact that, as long as the 1-flatness stabiliser conditions are satisfied, we can modify the 1-paths $\ell^\msf p_\mc P$ with respect to which the operator is defined. For instance, we have
\begin{equation}
    \cGateRelations{1} \,=\, \cGateRelations{2} \, .
\end{equation}
Since operators of the form
\begin{equation}
    \label{eq:commuteOneCharge}
    \cGateCommute{1} \q {\rm and} \q  \cGateCommute{2}  
\end{equation}
always commute, this guarantees that the membrane operator defined in eq.~\eqref{eq:exampleOneCharge} commutes with every edge operator in the interior of $\mc P$. Similarly, it follows from the commutation relations given in eq.~\eqref{eq:commutRelationsCSX} that operators of the form
\begin{equation}
    \cGateCommute{3} \q {\rm and} \q \cGateCommute{4}
\end{equation}
always commute, which in turn guarantees that the membrane operator defined in eq.~\eqref{eq:exampleOneCharge} commutes with every vertex operator in the interior of $\mc P$. More generally, the interiors of the membrane-like operators defined according to eq.~\eqref{eq:oneChargeOp} are 0- and 1-form gauge invariant.

Naturally, when picking $k = 0 \in \mathbb Z_3$, the 1-charge $\mc E_{\omega^k}$ reduces to the condensation defect $\mc E_1$ previously defined.
Notice however that since they appear at the boundary of topological membrane-like operators---in contrast to the condensation defect $\mc E_1$---the 1-charges $\mc E_\omega$ and $\mc E_{\bar \omega}$ necessarily have the topology of a circle, i.e. there is no domain wall between them and the trivial 1-charge $\mathbb 1$, nor there are domain walls between $\mc E_{\omega / \bar \omega}$ and $\mc E_1$. Moreover, there are only trivial domain walls between two 1-charges $\mc E_{\omega}$ or two 1-charges $\mc E_{\bar \omega}$, namely $1_{\mc E_{\omega}} : \mc E_\omega \to \mc E_\omega$ and $1_{\mc E_{\bar \omega}} : \mc E_{\bar \omega} \to \mc E_{\bar \omega}$. But there are non-trivial domain walls $n_{\omega} : \mc E_{\omega} \to \mc E_{\bar \omega}$ and $n_{\bar \omega} : \mc E_{\bar \omega} \to \mc E_{\omega}$. These are constructed as follows: Consider a rectangular 2-path $\mc P$, halve it for instance vertically, and apply operators \eqref{eq:oneChargeOp} so as to define a 1-charge $\mc E_{\omega}$ on the left-hand side and a 1-charge $\mc E_{\bar \omega}$ on the right-hand side. The domain wall itself then amounts to altering the condensation defect along $\partial \mc P$ by projecting the interfacial qubits on the side of the basepoint ${\rm bp}(\mc P)$ onto the spin-$\downarrow$ subspace. For instance, we have
\begin{equation}
    \label{eq:OneChargeMor}
    \OneChargeMor \, ,
\end{equation}
where we recall that the wiggly purple lines $\wigLine{2}$ indicate a qubit that is projected onto the spin-$\downarrow$ subspace. The domain wall $n_{\bar \omega} : \mc E_{\bar \omega} \to \mc E_{\omega}$ is defined similarly. Let us confirm that these domain walls are such that the resulting defects are well-defined 1-charges. More specifically, we need to confirm that only stabiliser conditions along the boundary of the 2-paths are violated. Consider for instance the two adjoining plaquettes in the middle of the operator considered in eq.~\eqref{eq:OneChargeMor}. In order to compute the commutation relation with the interstitial edge operator, we must first modify the 1-path $\ell^\msf p_\mc P$ associated with the plaquette on the left-hand side. The operator is modified as follows:
\begin{equation}
    \cGateRelations{3} \,=\, \cGateRelations{4}
\end{equation}
We can then invoke the commutation of operators of the form \eqref{eq:commuteOneCharge} to confirm that the edge stabiliser condition is indeed fulfilled. Similarly, we can confirm that all the edge stabiliser conditions in the interior of $\mc P$ are satisfied as well as all the vertex stabiliser constraints.

This concludes the enumeration of the boundary excitations of the higher gauge model. We summarise our findings so far in the table below:
\begin{align*}
	\setlength{\tabcolsep}{.4em}
	\begin{tabularx}{.96\columnwidth}{c|cccc}
         & $\mathbb 1$ & $\mc E_1$ & $\mc E_\omega$ & $\mc E_{\bar \omega}$         
        \\
        \midrule
        $\mathbb 1$ & $\mathbb 1 \xrightarrow{1_\mathbb 1,e} \mathbb 1$ & $\mathbb 1 \xrightarrow{\ldw} \mc E_1$ & $-$ & $-$
        \\
        $\mc E_1$ & $\mc E_1 \xrightarrow{\rdw} \mathbb 1$ & $\mc E_1 \xrightarrow{1_{\mc E_1},n} \mc E_1$ & $-$ & $-$ 
        \\
        $\mc E_\omega$ & $-$ & $-$ & $\mc E_\omega \xrightarrow{1_{\mc E_\omega}} \mc E_\omega$ & $\mc E_\omega \xrightarrow{n_{\omega}} \mc E_{\bar \omega}$
        \\
        $\mc E_{\bar \omega}$ & $-$ & $-$ & $\mc E_{\bar \omega} \xrightarrow{n_{\bar \omega}} \mc E_\omega$ & $\mc E_{\bar \omega} \xrightarrow{1_{\mc E_{\bar \omega}}} \mc E_{\bar \omega}$
	\end{tabularx} \, .
\end{align*}
In sec.~\ref{sec:TRep} we shall recover these 1-charges and the domain walls between them as the simple 2-representations of the 2-group $\mathbb G(2,3)$ and simple 1-intertwiners between them, respectively. 

As a final remark, let us comment on the physical interpretation of these defects as categorical charges. By definition, a ground state of the higher gauge model satisfies all the stabiliser conditions, i.e. it is in the $+1$ eigenspace of all the operators. In particular, it is invariant under 0- and 1-form gauge transformations, as generated by the vertex and edge operators, respectively. At the boundary, these ground state constraints enforce global 0- and 1-form symmetry conditions, respectively. The corresponding symmetry operators can be obtained by composing vertex and edge operators of the bulk Hamiltonian. In virtue of the commutation relation between vertex and edge operators, 0-form symmetry operators act on 1-form symmetry operators, resulting in a global symmetry encoded into the 2-group $\mathbb G(2,3)$. By definition such symmetry operators act trivially on the quasi-two-dimensional system associated with our choice of gapped boundary condition. In other words, this is the symmetry-preserving gapped boundary condition. In that context, we can interpret the boundary excitations derived in this section as categorical charges with respect to this global 2-group symmetry that is enforced by the bulk topological order.

\subsection{Excitation statistics}

\noindent
Let us now compute the fusion rules of the point-like domain wall excitations on the one hand and the fusion of the 1-charges on the other hand.\footnote{Although we refer to the `fusion' of domain walls, it is usually not encoded into a monoidal structure but rather a composition rule.} We begin with domain walls of the trivial 1-charge $\mathbb 1$. Trivially, we have $1_\mathbb 1 \circ 1_\mathbb 1 \simeq 1_\mathbb 1$ and $1_\mathbb 1 \circ e \simeq e \simeq e \circ 1_\mathbb 1$. Moreover, in virtue of $\sigma_Z^2 = {\rm id}$, we have $e \circ e \simeq 1_\mathbb 1$. Let us pursue with domain walls of the 1-charge $\mc E_1$. The only non-trivial fusion rule is associated with $n \circ n$. Recall that a domain wall $n$ may be interpreted as a violation of the 1-flatness stabiliser conditions surrounding an edge along the condensation defect. Bringing two such domain walls next to one another, the resulting excitation can be annihilated with a local operator:
\begin{align}
    \label{eq:condStringMono}
    \condStringMono{1} \simeq \condStringMono{2} \, ,
\end{align}
and thus $n \circ n \simeq 1_{\mc E_1}$. The remaining fusion rules involving the 1-charge $\mc E_1$ are $\ldw \circ \rdw$ and $\rdw \circ \ldw$. First of all, remark that, by construction, the output of these fusion processes must be domains walls $\mathbb 1 \to \mathbb 1$ and $\mc E_1 \to \mc E_1$, respectively. Focusing on the first scenario, let us bring together the domain walls $\rdw$ and $\ldw$ as follows:
\begin{align}
    \condStringComp{1} \, .
\end{align}
The qubit located in the middle of the horizontal edge in-between the domain walls $\rdw$ and $\ldw$ can either be $\uparrow$ or $\downarrow$. This gives rise to a decomposition of the Hilbert space:
\begin{align}
    \condStringComp{2} \; \oplus \; \condStringComp{3} \, .
\end{align}
In the first configuration, the qubit being projected onto the spin-$\uparrow$ subspace, the two condensation defects get connected. In the second configuration, we can think of the qubit being projected onto the spin-$\downarrow$ subspace as an alternation of the condensation defect as in eq.~\eqref{eq:altCondDefect}, or equivalently as the outcome of acting with a $\sigma_X$ operator on a short condensation defect $\mc E_1$ effectively causing a violation of the 1-flatness stabiliser conditions associated with the surrounding plaquettes, i.e.
\begin{align}
    \condStringComp{4} \; \oplus \; \condStringComp{5} \, .
\end{align}
Notice however that for this interpretation to be self-consistent, it is required to apply the same mechanism somewhere else on the condensation defect, or along a different condensation defect, as the resulting wave function would not be in the ground state subspace otherwise.
Putting everything together, we find $\rdw \circ \ldw \simeq 1_{\mc E_1} \oplus n$. Conversely, bringing together $\ldw$ and $\rdw$ yields a short condensation defect $\mc E_1$ that amounts to the local operator creating a pair of non-elementary point-like electric excitations decomposing as $1_\mathbb 1 \oplus e$, and therefore $\ldw \circ \rdw \simeq 1_\mathbb 1 \oplus e$.

The fusion rules computed so far coincide with those found on the smooth boundary of the (3+1)d toric code as computed in ref.~\cite{Kong:2020wmn}. Let us now consider the fusion of domain walls involving the 1-charges $\mc E_\omega$ and $\mc E_{\bar \omega}$. The only non-trivial fusion rules are those associated with $n_{\omega} \circ n_{\bar \omega}$ and $n_{\bar \omega} \circ n_{\omega}$. Given the obvious symmetry, it is sufficient to focus on the former. This situation arises when a `thin' membrane-like operator creating a 1-charge $\mc E_{\bar \omega}$ is inserted between two membrane-like operators creating two 1-charges $\mc E_{\omega}$, e.g.  
\begin{align}
    \OneChargeMorMono{1} \, .
\end{align}
Acting with vertex operators at the vertices where the domain walls meet has the effect of flipping the qubits projected onto the spin-$\downarrow$ subspace, thus annihilating the domain walls, while turning the 1-charge $\mc E_{\bar \omega}$ into $\mc E_\omega$ without introducing any magnetic excitations. This follows in particular from $\Gamma \Sigma_Z^\dagger = \Sigma_Z \Gamma$ as well as $(\sigma_X \otimes {\rm id}) \cSZ^\dagger = \cSZ (\sigma_X \otimes {\rm id})$. Therefore, the result is a single 1-charge $\mc E_{\omega}$. Therefore, we have $n_{\omega} \circ n_{\bar \omega} \simeq 1_{\mc E_\omega}$, and by symmetry we also find $n_{\bar \omega} \circ n_{\omega} = 1_{\mc E_{\bar \omega}}$. Note that, together, these fusions rules imply in particular that the 1-charges $\mc E_\omega$ and $\mc E_{\bar \omega}$ are isomorphic. We summarise the non-trivial fusion rules of domain walls we found below:
\begin{align}
    \begin{tikzcd}[ampersand replacement=\&, column sep=1.6em, row sep = 0em]
    	|[alias=A1]| \mathbb 1
    	\&
    	|[alias=B1]| \mathbb 1 
    	\&
    	|[alias=C1]| \mathbb 1 \,\simeq\, \mathbb 1
    	\&[1.2em]
    	|[alias=D1]| \mathbb 1 
    	\\
    	|[alias=A2]| \mc E_1
    	\&
    	|[alias=B2]| \mc E_1 
    	\&
    	|[alias=C2]| \mc E_1 \,\simeq\, \mc E_1
    	\&
    	|[alias=D2]| \mc E_1
    	\\
    	|[alias=A3]| \mathbb 1
    	\&
    	|[alias=B3]| \mc E_1 
    	\&
    	|[alias=C3]| \mathbb 1 \,\simeq\, \mathbb 1
    	\&
    	|[alias=D3]| \mathbb 1
    	\\
    	|[alias=A4]| \mc E_1
    	\&
    	|[alias=B4]| \mathbb 1
    	\&
    	|[alias=C4]| \mc E_1 \,\simeq\, \mc E_1
    	\&
    	|[alias=D4]| \mc E_1
    	\\
    	|[alias=A5]| \mc E_\omega
    	\&
    	|[alias=B5]| \mc E_{\bar \omega}
    	\&
    	|[alias=C5]| \mc E_\omega \,\simeq\, \mc E_\omega
    	\&
    	|[alias=D5]| \mc E_\omega
    	\\
    	|[alias=A6]| \mc E_{\bar \omega}
    	\&
    	|[alias=B6]| \mc E_\omega
    	\&
    	|[alias=C6]| \mc E_{\bar \omega} \,\simeq\, \mc E_{\bar \omega}
    	\&
    	|[alias=D6]| \mc E_{\bar \omega}
    	\arrow[rightarrow,from=A1,to=B1,"e"]
    	\arrow[rightarrow,from=B1,to=C1,"e"]
    	\arrow[rightarrow,from=C1,to=D1,"1_\mathbb 1"]
    	\arrow[rightarrow,from=A2,to=B2,"n"]
    	\arrow[rightarrow,from=B2,to=C2,"n"]
    	\arrow[rightarrow,from=C2,to=D2,"1_{\mc E_1}"]
    	\arrow[rightarrow,from=A3,to=B3,"\ldw"]
    	\arrow[rightarrow,from=B3,to=C3,"\rdw"]
    	\arrow[rightarrow,from=C3,to=D3,"1_\mathbb 1 \oplus e"]
    	\arrow[rightarrow,from=A4,to=B4,"\rdw"]
    	\arrow[rightarrow,from=B4,to=C4,"\ldw"]
    	\arrow[rightarrow,from=C4,to=D4,"1_{\mc E_1} \oplus n"]
    	\arrow[rightarrow,from=A5,to=B5,"n_\omega"]
    	\arrow[rightarrow,from=B5,to=C5,"n_{\bar \omega}"]
    	\arrow[rightarrow,from=C5,to=D5,"1_{\mc E_\omega}"]
    	\arrow[rightarrow,from=A6,to=B6,"n_{\bar \omega}"]
    	\arrow[rightarrow,from=B6,to=C6,"n_\omega"]
    	\arrow[rightarrow,from=C6,to=D6,"1_{\mc E_{\bar \omega}}"]
    \end{tikzcd} \, .
\end{align}

\bigskip \noindent
Let us now consider the fusion of the 1-charges themselves. First of all, we trivially have
\begin{equation}
    \mathbb 1 \boxtimes \mc E_\eta \cong  \mc E_\eta \cong \mc E_{\eta} \boxtimes \mathbb 1 \, ,
\end{equation}
where $\eta \in \{1,\omega,\bar \omega\}$ labels a character of $\mathbb Z_3$. Non-trivial fusion rules are those of the form $\mc E_\eta \boxtimes \mc E_\mu$, where $\eta,\mu \in \{1,\omega,\bar \omega\}$. For concreteness, let us focus on $\mc E_\omega \boxtimes \mc E_\omega$. Microscopically, such a fusion amounts to bringing the loop-like defects very close to one another and consider the resulting system as a single (typically non-elementary) excitation. As such this fusion does not strictly take place on the boundary, rather we should think of one of the 1-charges as coming from the bulk, e.g. 
\begin{equation}
    \label{eq:OneChargeMonoidal}
    \OneChargeMonoidal{1} \, . 
\end{equation}
The task is now to determine which excitation this system corresponds to.
The qubit degrees of freedom between the two 1-charges can a priori be $\uparrow$ or $\downarrow$, but it follows from the plaquette stabiliser conditions being satisfied that these interstitial qubits must all be in the same state. The previous configuration is thus equivalent to
\begin{equation*}
    \OneChargeMonoidal{2} \; \boxplus \;  \OneChargeMonoidal{3} \, .
\end{equation*}
On the one hand, we can think of qubits projected onto the spin-$\uparrow$ subspace as corresponding to the presence of condensation defects $\mc E_1$, and can thus be represented as wiggly red lines.
On the other hand, we can think of the qubits projected onto the spin-$\downarrow$ subspace as the result of acting with $\sigma_X$ operators on qubits in the $\uparrow$ state, causing violations of 1-flatness stabiliser conditions associated with the surrounding plaquettes. But recall that topological defects are defined up to genuine line-like operators. It turns out the resulting magnetic excitations may be annihilated by acting with genuine line-like operators on the boundary. In order to appreciate the fact that the required operators are indeed genuinely line-like, we need to consider larger 1-charges following the pattern described in eq.~\eqref{eq:exampleOneCharge}. Given a defect configuration of 1-charges and magnetic excitations, we consider genuine line-like operators of the form
\begin{equation}
    \OneChargeMonoidalCancel \, .
\end{equation}
Although the $\sigma_X$ operators are sufficient to annihilate the 1-fluxes arising from the violations of the 1-flatness stabiliser conditions, the resulting configuration would not commute with some of the cube operators---giving rise to 2-fluxes---if not for the addition of the $\Gamma$ operators. 
It follows from the commutation relations given in eq.~\eqref{eq:commutRelationsCSX} that the effect of such an operator---in addition to annihilating the magnetic excitations---is to turn the $\cSZ$ operators at the front into $\cSZ^\dagger$ operators.
Applying a smaller version of this operator, we thus find that the configuration depicted in eq.~\eqref{eq:OneChargeMonoidal} ends up being equivalent to
\begin{equation*}
    \OneChargeMonoidal{4} \boxplus  \OneChargeMonoidal{5} \, .
\end{equation*}
Since $\Sigma_Z^2 = \Sigma_Z^\dagger$, the first term in this decomposition amounts to a single `thick' 1-charge $\mc E_{\bar \omega}$, whereas the second term amounts to a single 1-charge $\mc E_{1}$. Putting everything together, we find the fusion rules $\mc E_{\omega} \boxtimes \mc E_{\omega} \cong \mc E_{\bar \omega} \boxplus \mc E_{1}$, and more generally
\begin{equation}
    \label{eq:fusion1Charges}
    \mc E_\eta \boxtimes \mc E_\mu 
    \cong 
    \mc E_{\eta \mu} \boxplus \, \mc E_{\eta \bar \mu}  \, ,
\end{equation}
for any $\eta,\mu \in \{1,\omega, \bar \omega\}$, where we think of $\mc E_{\mu}$ as being the 1-charge at the front. We recover in particular the fusion rules $\mc E_{1} \boxtimes \mc E_{1} \cong \mc E_{1} \boxplus \mc E_{1}$ of $\mathbb Z_2$ condensation defects \cite{PhysRevB.96.045136,Kong:2020wmn,Roumpedakis:2022aik}. 

Let us conclude this enumeration of the boundary excitations and their statistics with a couple of representation theoretic comments: We have already mentioned that a short string-like condensation defect $\mc E_1$ spanning over a single edge of the lattice amounts to the local operator creating a superposition $1_\mathbb 1 \oplus e$ of pairs of 0-charges. In the same vein, the \emph{tube algebra} analysis carried out in \cite{Bullivant:2019fmk,Bullivant:2019tbp} predicts that a \emph{small} loop-like electric excitation, to which a point-like electric excitation may be attached, are labelled by irreducible representations of $\mathbb D_6 \simeq \mathbb Z_2 \ltimes \mathbb Z_3$. Recall that $\mathbb D_6$ possesses three irreducible representations, namely $\{\ub 0, \ub 1, \ub 2\}$ referred to as the trivial, sign and standard representation. In that context, we can identify the 1-charges $\mc E_1$, $\mc E_\omega$ and $\mc E_{\bar \omega}$ studied in this section as being labelled by the induced representations ${\rm Ind}_{\mathbb Z_3}^{\mathbb D_6}(1) \simeq \ub 0 \oplus \ub 1$, ${\rm Ind}_{\mathbb Z_3}^{\mathbb D_6}(\omega) \simeq \ub 2$ and ${\rm Ind}_{\mathbb Z_3}^{\mathbb D_6}(\bar \omega) \simeq \ub 2$, respectively. Notice furthermore that this identification is compatible with the fusion rules \eqref{eq:fusion1Charges} in virtue of $\ub 2 \otimes \ub 2 \simeq (\ub 0 \oplus \ub 1) \oplus \ub 2$.

\section{Categorical representations\label{sec:TRep}}

\noindent
\emph{We show in this section how the categorical charges defined above as well as their statistics are organised into the monoidal bicategory of 2-representations of the 2-group $\mathbb G(2,3)$.}

\subsection{Preliminaries}

\noindent
In order to introduce the notion of 2-group 2-representation employed in this manuscript, let us first review the category theoretical definition of a group representation. We invite the reader to consult \cite{etingof2016tensor} for basic definitions. Every finite group $G$ can be thought as a one-object \emph{groupoid} referred to as the \emph{delooping} $\msf B G$ of $G$. It is the \emph{category} with unique object $\bul$ and hom-set $\Hom(\msf B G) = G \ni \bul \xrightarrow{g} \bul$ such that the composition rule is provided by multiplication in $G$, i.e. 
\begin{equation}
    \bul \xrightarrow{g_1} \bul \xrightarrow{g_2} \bul = \bul \xrightarrow{g_1g_2} \bul \, , \q \forall \, g_1,g_2 \in G \, .
\end{equation}
A representation of $G$ is then defined as a \emph{functor} from $\msf B G$ to $\Vect$, where $\Vect$ is the category of complex vector spaces and linear maps:
\begin{align}
    \arraycolsep=1.4pt
    \begin{array}{ccccl}
        \rho  & : & \msf B G & \to & \; \Vect
        \\[.4em]
        & : & \bul  & \mapsto & \; \rho(\bul) =: V
        \\
        & : & \bul \xrightarrow{g} \bul & \mapsto & \; V \xrightarrow{\rho(g)} V \in {\rm End}_\mathbb C(V)  
    \end{array} \, ,
\end{align}
i.e. it assigns a vector space $V$ to the unique object $\bul$ in $\msf B G$, and a linear map $\rho(g)$ in ${\rm End}_{\mathbb C}(V)$ to every morphism $\bul \xrightarrow{g}\bul$ such that $\rho(g_1 g_2)= \rho(g_1) \circ \rho(g_2)$ for any $g_1, g_2 \in G$. In the same vein, given two representations $\rho$ and $\sigma$, an intertwiner $\vartheta : \rho \to \sigma$ between them is defined as a \emph{natural transformation} between the corresponding functors, i.e. the assignment of a morphism $\vartheta_{\bulScr} : \rho(\bul) \to \sigma(\bul)$ to the unique object $\bul$ in $\msf B G$ such that $\rho(g) \circ \vartheta_{\bulScr} = \vartheta_{\bulScr} \circ \sigma(g)$ for every $g \in G$. Putting everything together we obtain that the category $\Rep(G)$ of representations and intertwiners can be defined as the \emph{functor category} $\Fun(\msf B G, \Vect)$.

\bigskip \noindent
Let us now go up one level of abstraction so as to enter the realm of 2-categories, where, in addition to objects and (1-)morphisms, there are 2-morphisms between 1-morphisms. Given two objects $A,B \in \Ob(\mc C)$ in a 2-category $\mc C$, we denote by $\HomC_{\mc C}(A,B)$ the category of morphisms (hom-category) from $A$ to $B$. As customary, objects and morphisms in hom-categories shall be referred to as 1- and
2-morphisms in $\mc C$, respectively. Given $A,B,C \in \Ob(\mc C)$, the \emph{horizontal} composition functor of 1-
and 2-morphisms is notated via $\circ : \HomC_\mc C(A, B) \times \HomC_\mc C(B, C) \to \HomC_\mc C(A, C)$, and the corresponding
identity 1- and 2-morphisms via $1_A$ and $1_{1_A}$, respectively. The \emph{vertical} composition of 2-morphisms within a hom-category $\HomC_\mc C(A,B)$ is denoted by $\cdot$ and the corresponding identity 1-morphisms by $1_f$ for any $f: A \to B$. Furthermore, we recall that a \emph{bicategory} is a 2-category for which the associativity of the horizontal composition is weakened by a natural 2-isomorphisms referred to as the \emph{1-associator}.

Given a 2-group $\mathbb G$ associated with a crossed module of the form $(G,H,\mathbb 1_G : H \to \{\mathbb 1_G\}, \act)$ with $H$ abelian,\footnote{Crossed modules with arbitrary group homomorphisms $\partial : G \to H$ with $H$ non-abelian can be treated in the exact same way, but since we ultimately focus on $\mathbb G(2,3)$ for which $\partial$ is trivial, we have no need dealing with the extra technicalities.} its delooping $\msf B \mathbb G$ is a 2-groupoid consisting of a single object $\bul$ with hom-category $\HomC_{\msf B \mathbb G}(\bul,\bul) = \mathbb G$ such that horizontal and vertical compositions are provided by the monoidal structure in $\mathbb G$ and the composition of (1-)morphisms in $\mathbb G$, respectively \cite{2003math......7200B}. More concretely, 1-morphisms are of the form $\bul \xrightarrow{g} \bul$, with $g \in G$, and compose via group multiplication in $G$, whereas 2-morphisms are of the form
\begin{equation}
	\begin{tikzcd}[ampersand replacement=\&, column sep=2.5em, row sep=2.5em,line cap=butt]
        |[alias=A]| \bul \&
        |[alias=B]| \bul
        \arrow[from=A,to=B,"g",bend left=50,""{name=X,below}]
        \arrow[from=A,to=B,"g"',bend right=50,""{name=Y,above}]
        \arrow[Rightarrow,from=X,to=Y,"h"]
    \end{tikzcd} \, , \q \forall \, g \in G, h \in H \, .
\end{equation}
Vertical composition of 2-morphisms is defined via group multiplication in $H$ as
\begin{equation}
	\begin{tikzcd}[ampersand replacement=\&, column sep=2.5em, row sep=2.5em,line cap=butt]
        |[alias=A]| \bul \&
        |[alias=B]| \bul
        \arrow[from=A,to=B,"g",bend left=80,""{name=X,below}]
        \arrow[from=A,to=B,""{name=Y,above},""{name=Y1,above},""{name=Y2,below}]
        \arrow[from=A,to=B,"g"',bend right=80,""{name=Z,above}]
        \arrow[Rightarrow,from=X,to=Y1,"h_1"]
        \arrow[Rightarrow,from=Y2,to=Z,"h_2"]
    \end{tikzcd} 
    =
	\begin{tikzcd}[ampersand replacement=\&, column sep=2.5em, row sep=2.5em,line cap=butt]
        |[alias=A]| \bul \&
        |[alias=B]| \bul
        \arrow[from=A,to=B,"g",bend left=50,""{name=X,below}]
        \arrow[from=A,to=B,"g"',bend right=50,""{name=Y,above}]
        \arrow[Rightarrow,from=X,to=Y,"\tilde h"]
    \end{tikzcd}
\end{equation}
with $\tilde h := h_1+h_2$ for any $g \in G$ and $h_1,h_2 \in H$,
whereas horizontal composition of 2-morphisms is provided by group multiplication in $G \ltimes H$ as
\begin{equation}
	\begin{tikzcd}[ampersand replacement=\&, column sep=2.5em, row sep=2.5em,line cap=butt]
        |[alias=A]| \bul \&
        |[alias=B]| \bul \&
        |[alias=C]| \bul 
        \arrow[from=A,to=B,"g_1",bend left=50,""{name=X1,below}]
        \arrow[from=A,to=B,"g_1"',bend right=50,""{name=Y1,above}]
        \arrow[Rightarrow,from=X1,to=Y1,"h_1"]
        \arrow[from=B,to=C,"g_2",bend left=50,""{name=X2,below}]
        \arrow[from=B,to=C,"g_2"',bend right=50,""{name=Y2,above}]
        \arrow[Rightarrow,from=X2,to=Y2,"h_2"]
    \end{tikzcd}
    =
	\begin{tikzcd}[ampersand replacement=\&, column sep=2.5em, row sep=2.5em,line cap=butt]
        |[alias=A]| \bul \&
        |[alias=B]| \bul
        \arrow[from=A,to=B,"g_1g_2",bend left=50,""{name=X,below}]
        \arrow[from=A,to=B,"g_1g_2"',bend right=50,""{name=Y,above}]
        \arrow[Rightarrow,from=X,to=Y,"\tilde h"]
    \end{tikzcd} 
\end{equation}
with $\tilde h := h_1+ (g_1 \act h_2)$ for any $g_1,g_2 \in G$ and $h_1,h_2 \in H$.

In order to introduce a notion of 2-representations, we require a \emph{categorification} of the notion of vector spaces. We denote by $\TVect$ the 2-category of finite semi-simple linear categories,  linear functors and natural transformations. As customary, we shall refer to objects in $\TVect$ as \emph{2-vector spaces}. This bicategory can be further equipped with a monoidal structure via the Deligne tensor product $\boxtimes$ of abelian categories. Note that $\TVect$ has a unique simple object, namely $\Vect$.

\bigskip \noindent
Given the above, we define the bicategory $\TRep(\mathbb G)$ of 2-representations, 1-interwiners and 2-intertwiners of a 2-group $\mathbb G$ as the bicategory $\TFun(\msf B \mathbb G, \TVect)$ of \emph{pseudofunctors}, \emph{pseudonatural transformations} and \emph{modifications}. Instead of providing the general definition of these notions, let us immediately unpack the bicategory $\TFun(\msf B \mathbb G, \TVect)$. 

A 2-representation $\rho$ of $\mathbb G$ is a map
\begin{align}
    \arraycolsep=1.4pt
    \begin{array}{ccccl}
        \rho  & : & \msf B \mathbb G & \to & \; \TVect
        \\[.4em]
        & : & \bul  & \mapsto & \; \rho(\bul) =: \mc V
        \\[.3em]
        & : & \bul \xrightarrow{\, g\, } \bul & \mapsto & \; \mc V \xrightarrow{\rho(g)} \mc V \in \EndC(\mc V)
        \\
        & : & \!
    	\begin{tikzcd}[ampersand replacement=\&, column sep=2.5em, row sep=2.5em,line cap=butt]
            |[alias=A]| \bul \&
            |[alias=B]| \bul
            \arrow[from=A,to=B,"g",bend left=50,""{name=X,below}]
            \arrow[from=A,to=B,"g"',bend right=50,""{name=Y,above}]
            \arrow[Rightarrow,from=X,to=Y,"h"]
        \end{tikzcd} \!
        & \mapsto & \!
    	\begin{tikzcd}[ampersand replacement=\&, column sep=2.5em, row sep=2.5em,line cap=butt]
            |[alias=A]| \mc V \&
            |[alias=B]| \mc V
            \arrow[from=A,to=B,"\rho(g)",bend left=50,""{name=X,below}]
            \arrow[from=A,to=B,"\rho(g)"',bend right=50,""{name=Y,above}]
            \arrow[Rightarrow,from=X,to=Y,"\rho(h)"]
        \end{tikzcd} \!\! \in \End_{\EndC(\mc V)}(\rho(g))
    \end{array} \, ,
\end{align}
where $\mc V$ is a finite linear semisimple (1-)category, $\rho(g)$ are endofunctors of $\mc V$ and $\rho(h): \rho(g) \Rightarrow \rho(g)$ are natural transformations. 
Furthermore, $\rho$ is required to strictly preserve vertical and horizontal compositions of 2-morphisms as well as to preserve composition of 1-morphisms up to natural isomorphisms. In other words, we have 
\begin{align}
    \label{eq:LinearTMorV}
    \rho(h_1) \cdot \rho(h_2) &= \rho(h_1 \cdot h_2)
    \\
    \label{eq:LinearTMorH}
    \rho(h_1) \circ \rho(h_2) &= \rho(h_1 \circ h_2)
     \, ,
\end{align}
for any $h_1: g_1 \Rightarrow g_1$, $h_2 : g_2 \Rightarrow g_2$ labelled by $h_1,h_2 \in H$, and $\rho$ assigns to every pair of 1-morphisms labelled by $g_1,g_2 \in G$ a natural 2-isomorphism
\begin{equation}
    \label{eq:ModAssociator}
    \rho_{g_1,g_2}: \rho(g_1) \circ \rho(g_2) \xRightarrow{\sim} \rho(g_1g_2) \, ,
\end{equation}
which is required to fulfil
\begin{equation}
    [1_{\rho_(g_1)} \circ \rho_{g_2,g_3}] \cdot \rho_{g_1,g_2g_3} = [\rho_{g_1,g_2} \circ 1_{\rho(g_3)}] \cdot \rho_{g_1g_2,g_3}  \, ,  
\end{equation}
for any $g_1,g_2,g_3 \in G$.

Given two 2-representations $\rho$ and $\sigma$ of $\mathbb G$, a 1-intertwiner $\vartheta : \rho \to \sigma$ between them consists of an object $\vartheta_{\bulScr}$ in $\Fun(\rho(\bul), \sigma(\bul))$ together with a collection of natural 2-isomorphisms $\vartheta_{-}$ defined via
\begin{equation}
    \vartheta_g : \rho(g) \circ \vartheta_{\bulScr} \xRightarrow{\sim} \vartheta_{\bulScr} \circ \sigma(g) \, , \q g \in G \, .
\end{equation}
such that $\vartheta_{\mathbb 1_G} = 1_{\vartheta_{\bulScr}}$. This collection of 2-isomorphisms is subject to a coherence relation ensuring compatibility with the horizontal composition of 1-morphisms in $\mathbb G$, namely
\begin{align}
    \label{eq:ModFun}
    &[\rho_{g_1,g_2} \circ 1_{\vartheta_{\bulScr}}] \cdot \vartheta_{g_1g_2} 
    \\
    \nn
    & \q = [1_{\rho(g_1)} \circ \vartheta_{g_2}] \cdot [\vartheta_{g_1} \circ 1_{\sigma(g_2)}] \cdot [1_{\vartheta_{\bulScr}} \circ \sigma_{g_1,g_2}]
\end{align}
for any $g_1,g_2 \in G$. Moreover, the naturality condition stipulates that
\begin{equation}
    \label{eq:NatInt}
    [\rho(h) \circ 1_{\vartheta_{\bulScr}}] \cdot \vartheta_g = \vartheta_g \cdot [1_{\vartheta_{\bulScr}} \circ \sigma(h)]
\end{equation}
for any 2-morphism $h : g \Rightarrow g$ in $\mathbb G$. 

Finally, given two 1-intertwiners $\vartheta,\tilde \vartheta : \rho \to \sigma$ between two 2-representations $\rho$ and $\sigma$ of $\mathbb G$, a 2-intertwiner $\Theta : \vartheta \Rightarrow \tilde \vartheta$ between them consists of a natural transformation $\Theta_{\bulScr} : \vartheta_{\bulScr} \Rightarrow \tilde \vartheta_{\bulScr}$ in $\Fun(\rho(\bul),\sigma(\bul))$ such that
\begin{equation}
    \vartheta_g \cdot [\Theta_{\bulScr} \circ 1_{\sigma(g)}] = [1_{\rho(g)} \circ \Theta_{\bulScr}] \cdot \tilde \vartheta_g
\end{equation}
is fulfilled for any $g \in G$.

\bigskip \noindent
The definitions above are still somewhat terse so let us unfold them further before specialising to $\mathbb G(2,3)$. Let us begin with 2-representations. A finite semisimple 1-category $\mc V$ together with endofunctors $\rho(g): \mc V \to \mc V$, for every $g \in G$, and natural 2-isomorphisms $\rho_{g_1,g_2} : \rho(g_1) \circ \rho(g_2) \Rightarrow \rho(g_1g_2)$ in $\TVect$ satisfying the coherence relation given in eq.~\eqref{eq:ModAssociator} define a right \emph{module} category over the category $\Vect_G$ of $G$-graded vector spaces. Recall that simple objects in $\Vect_G$ are the one-dimensional vector spaces $\mathbb C_g$, for every $g \in G$, such that $\mathbb C_{g_1} \otimes \mathbb C_{g_2} \simeq \mathbb C_{g_2}$ and $\Hom_{\Vect_G}(\mathbb C_{g_1},\mathbb C_{g_2}) = \delta_{g_1,g_2} \mathbb C$. Introducing the notations $\cat : \mc V \times \Vect_G \to \mc V$, whereby $M \cat \mathbb C_g := \rho(g)(M)$ for every $M \in \Ob(\mc V)$, and $\alpha^{\cat}_{M,\mathbb C_{g_1}, \mathbb C_{g_2}} := (\rho_{g_1,g_2})_M$, it follows from the 2-cocycle condition satisfied by $\rho_{g_1,g_2}$ that
\begin{align}
    & \alpha^{\cat}_{M \cat \mathbb C_{g_1},\mathbb C_{g_2},\mathbb C_{g_3}} 
    \circ \alpha^{\cat}_{M,\mathbb C_{g_1},\mathbb C_{g_2g_3}}
    \\
    \nn
    & \q =
    (\alpha^{\cat}_{M,\mathbb C_{g_1},\mathbb C_{g_2}} \cat 1_{\mathbb C_{g_3}})
    \circ
    \alpha^{\cat}_{M,\mathbb C_{g_1g_2},\mathbb C_{g_3}}
    \circ 1_{M \cat \mathbb C_{g_1g_2g_3}}
\end{align}
holds for all $g_1,g_2,g_3 \in G$ and $M \in \Ob(\mc V)$, and thus $\alpha^{\cat} : (- \cat -) \cat - \to - \cat (- \otimes -)$ defines a \emph{module associator} for $\mc V$ with respect to $\cat$. Putting everything together, the triple $(\mc V, \cat, \alpha^{\cat})$ does define a \emph{right} module category over $\Vect_G$. In virtue of $\rho(h)$ defining a natural transformation between $\rho(g)$ and itself, together with the compatibility condition eq.~\eqref{eq:LinearTMorV}, we find that $\rho$ further assigns to every object $M \in \Ob(\mc V)$ a representation $\rho(-)_M : H \to \End_{\mc V}(\rho(g)(M))$ of $H$ on $\rho(g)(M) \in \Ob(\mc V)$ for any $g \in G$, and to every 1-morphism in $\mc V$ an intertwiner between the corresponding representations. 
The compatibility condition \eqref{eq:LinearTMorH} finally requires that
\begin{equation}
    \label{eq:LinearTMorHb}
    \rho(g \act -)_M = \rho(-)_{\rho(g)(M)}    
\end{equation} 
for every $g \in G$ and $M \in \Ob(\mc V)$.

Let us now examine further 1-intertwiners at the light of this interpretation of 2-representations. Given two 2-representations $\rho$ and $\sigma$ of $\mathbb G$ with underlying $\Vect_G$-module categories $(\mc V,\cat,\alpha^{\cat})$ and $(\mc W,\catb,\alpha^{\catb})$, respectively, an object $\vartheta_{\bulScr}$ in $\Fun(\mc V,\mc W)$ together with a collection of natural 2-isomorphisms $\vartheta_g : \rho(g) \circ \vartheta_{\bulScr} \Rightarrow \vartheta_{\bulScr} \circ \sigma(g)$ satisfying eq.~\eqref{eq:ModFun} defines a $\Vect_G$-\emph{module functor} from $\mc V$ to $\mc W$. Introducing the notation, $\omega_{M,\mathbb C_g} := (\vartheta_g)_M$, it follows from eq.~\eqref{eq:ModFun} that
\begin{align}
    &\vartheta_{\bulScr}(\alpha^{\cat}_{M,\mathbb C_{g_1},\mathbb C_{g_2}})
    \circ \omega_{M,\mathbb C_{g_1g_2}}
    \\
    \nn
    & \q =
    \omega_{M \cat \mathbb C_{g_1},\mathbb C_{g_2}} 
    \circ (\omega_{M,\mathbb C_{g_1}} \catb 1_{\mathbb C_{g_2}})
    \circ \alpha^{\catb}_{\vartheta_{\bulScr}(M),\mathbb C_{g_1},\mathbb C_{g_2}}
\end{align}
holds for every $g_1,g_2 \in G$ and $M \in \Ob(\mc V)$, and thus the pair $(\vartheta_{\bulScr},\omega)$ with $\omega : \vartheta_{\bulScr}(- \cat -) \to \vartheta_{\bulScr}(-)\catb -$ does define a $\Vect_G$-module functor. The category of such module functors over $\Vect_G$ is denoted by $\Fun_{\Vect_G}(\mc V,\mc W)$. The naturality condition \eqref{eq:NatInt} further stipulates that 
\begin{equation}
    \label{eq:NatIntb}
    \vartheta_{\bulScr}(\rho(h)_M) \circ \omega_{M,\mathbb C_g}
    =
    \omega_{M,\mathbb C_g} \circ \sigma(h)_{\vartheta_{\bulScr}(M)}
\end{equation}
must hold for every $h: g \Rightarrow g$ and $M \in \Ob(\mc V)$, and thus not all objects in $\Fun_{\Vect_G}(\mc V,\mc W)$ necessarily provide a 1-intertwiner between the corresponding 2-representations.

Before computing explicitly the bicategory $\TRep(\mathbb G(2,3))$, let us remark that in the limiting case where we forget about the group $\mathbb Z_3$, which amounts to treating the group $\Zt$ as a 2-group, the construction presented in this section yields the monoidal bicategory $\TRep(\Zt)$ studied by Kong et al. in \cite{Kong:2020wmn} in the context of the (3+1)d toric code. 

\subsection{2-group 2-representations}

\noindent
Let us now apply the definitions spelt out above to the case of $\mathbb G(2,3)$ beginning with the computation of all the (simple) 2-representations.\footnote{We are referring here to the notion of semi-simplicity in the sense of Douglas and Reutter \cite{douglas2018fusion}.} Note that a similar derivation was carried out by Barrett and Mackaay in \cite{Barrett:2004zb} but their definition of $\TRep(\mathbb G(2,3))$ differs a little it from the one employed in this manuscript.

We established earlier that a 2-representation of $\mathbb G(2,3)$ is labelled---amongst other things---by a choice of right module category over $\Vect_{\Zt}$. There are only two indecomposable $\Vect_{\Zt}$-module categories, namely $\Vect_{\Zt}$ itself via the monoidal product in $\Vect_{\Zt}$ and $\Vect$ via the forgetful functor $\Vect_{\Zt} \to \Vect$ \cite{2002math......2130O}. Notice that both module categories have module associators $\alpha^{\cat}$ given by the identity morphisms. Let us treat these two cases separately:

\bigskip\noindent
$*$ Let $\rho$ be a 2-representation of $\mathbb G(2,3)$ whose underlying $\Vect_{\Zt}$-module category $(\mc V:= \rho(\bul),\cat,\alpha^{\cat})$ is $\Vect$. The 2-representation $\rho$ assigns to the unique simple object in $\Vect$, namely $\mathbb C$, a representation $\rho(-)_\mathbb C : \ZT \to \End_{\mc V}(\rho(g)(\mathbb C))$ for any $g \in \Zt$. Since $\rho(g)(\mathbb C) \simeq \mathbb C$, $\rho(-)_\mathbb C$ amounts to a \emph{character} on $\mathbb C$. This one-dimensional representation of $\ZT$ is further required to fulfil $\rho(g \act -)_\mathbb C = \rho(-)_{\mathbb C}$ for every $g \in \Zt$, forcing $\rho(-)_\mathbb C$ to be the trivial representation of $\ZT$. We notate the resulting 2-representation of $\mathbb G(2,3)$ via $\mathbb 1$, which was referred to as the trivial 1-charge in the previous section. 

\medskip \noindent
$*$ Let $\rho$ be a 2-representation of $\mathbb G(2,3)$ whose underlying $\Vect_{\Zt}$-module category $(\mc V,\cat,\alpha^{\cat})$ is $\Vect_{\Zt}$. The 2-representation $\rho$ assigns to the simple objects $\mathbb C_{+1}$ and $\mathbb C_{-1}$ in $\Vect_{\Zt}$ representations $\rho(-)_{\mathbb C_{\pm 1}} : \ZT \to \End_{\mc V}(\rho(g)(\mathbb C_{\pm 1}))$ for any $g \in \Zt$. Since $\rho(g)(\mathbb C_{\pm 1}) = \mathbb C_{\pm 1} \cat \mathbb C_g \simeq \mathbb C_{\pm g}$ and $\End_{\mc V}(\mathbb C_{\pm g}) \simeq \mathbb C$, $\rho(-)_{\mathbb C_{\pm 1}}$ are $\mathbb C$-valued characters of $\ZT$. These characters are further required to fulfill eq.~\eqref{eq:LinearTMorHb}, i.e.
$\rho(g \act -)_{\mathbb C_{\pm 1}} = \rho(-)_{\mathbb C_{\pm 1} \cat \mathbb C_{g}} = \rho(-)_{\mathbb C_{\pm g}}$. Choosing $g = -1 \in \Zt$, this imposes that $\rho(h^{-1})_{\mathbb C_{+1}} = \rho(h)_{\mathbb C_{-1}}$ for every $h \in \ZT$ and thus $\rho(-)_{\mathbb C_{-1}} = \rho(-)^{-1}_{\mathbb C_{+1}}$. We thus obtain three distinct 2-representations labelled by a complex character of $\mathbb Z_3$. These correspond to the non-trivial 1-charges $\mc E_1$, $\mc E_{\omega}$ and $\mc E_{\bar \omega}$ derived in the previous section. Notice that condition \eqref{eq:LinearTMorHb} precisely translates the fact that we required $\cSZ$ gates in order to construct membrane operators creating 1-charges along their boundaries.

\subsection{1-intertwiners}

\noindent
We pursue our analysis with the derivation of the 1-intertwiners between the four 2-representations defined above:

\bigskip \noindent 
$\mathbb 1 \to \mathbb 1$: A 1-intertwiner between the 2-representation $\mathbb 1$ and itself is specified---amongst other things---by a choice of $\Vect_{\Zt}$-module endofunctors of $\Vect$, i.e. an object in $\Fun_{\Vect_{\Zt}}(\Vect,\Vect)$. But we can easily show that the category $\Fun_{\Vect_{\Zt}}(\Vect,\Vect)$ is equivalent as a monoidal category to $\Rep(\Zt)$ \cite{etingof2016tensor}. Such a $\Vect_{\Zt}$-module functor is further required to fulfil eq.~\eqref{eq:NatIntb}, but it turns out to be trivial in that case given that $\rho(-)_{\mathbb C}$ corresponds to the trivial representation of $\ZT$. We thus identify two simple 1-intertwiners $\mathbb 1 \to \mathbb 1$ denoted by $1_\mathbb 1$ and $e$, associated with the two simple objects in $\Rep(\mathbb Z_2)$. These correspond to the 0-charges constructed in the previous section, thought as domain walls between two trivial 1-charges.

\medskip \noindent
$\mc E_1 \to \mc E_1$: Such a 1-intertwiner is now specified by a choice of $\Vect_{\Zt}$-module endofunctors of $\Vect_{\Zt}$, whereby $\Fun_{\Vect_{\Zt}}(\Vect_{\Zt},\Vect_{\Zt}) \cong \Vect_{\Zt}$. Similarly to the previous scenario, since the 2-representation $\mc E_1$ assigns the trivial character of $\mathbb Z_3$ to both simple objects in $\Vect_{\Zt}$, the naturality condition \eqref{eq:NatIntb} happens to be trivially satisfied. We thus identify two simple 1-intertwiners $\mc E_1 \to \mc E_1$ denotes by $1_{\mc E_1}$ and $n$, associated with the two simple objects in $\Vect_{\Zt}$. These correspond to the two types of domain walls between condensation defects constructed in the previous section.  

\medskip \noindent
$\mc E_{\omega / \bar \omega} \to \mc E_{\omega / \bar \omega}$: As for the previous case, such a 1-intertwiner is specified by an object in $\Vect_{\Zt}$ thought as a module endofunctor of $\Vect_{\Zt}$ over itself. But the naturality condition \eqref{eq:NatIntb} is not trivial anymore. There are two simple objects in $\Vect_{\Zt}$, namely $\mathbb C_{+1}$ and $\mathbb C_{-1}$, and the corresponding module functors are given by $ \mathbb C_{+1} \otimes -$ and $\mathbb C_{-1} \otimes -$, respectively, with the isomorphisms $\omega$ being identity morphisms. Focusing on $\mathbb C_{-1} \otimes -$, the naturality condition stipulates that we must have $1_{\mathbb C_{-1}} \otimes \rho(h)_{\mathbb C_{\pm 1}} = \rho(h)_{\mathbb C_{-1} \otimes \mathbb C_{\pm 1}}$, i.e. $\rho(h)_{\mathbb C_{\pm 1}}  = \rho(h)_{\mathbb C_{\mp 1}}$, which is not satisfied whenever $h \neq 0 \in \ZT$ given that $\rho(h)_{\mathbb C_{\mp 1}} = \rho(h^{-1})_{\mathbb C_{\pm 1}}$. On other hand, the naturality condition is always satisfied for $\mathbb C_{+1} \otimes -$. It follows that there is a unique simple 1-intertwiner $\mc E_{\omega / \bar \omega} \to \mc E_{\omega / \bar \omega}$, which we notate via $1_{\mc E_{\omega / \bar \omega}}$.

\medskip \noindent
$\mc E_{\omega / \bar \omega} \to \mc E_{\bar \omega / \omega}$: This case is very similar to the previous one with the difference that the module endofunctor $\mathbb C_{+1} \otimes -$ over itself is the one that does not satisfy the naturality condition. Indeed, the naturality condition stipulates that $1_{\mathbb C_{+1}} \otimes \rho(h)_{\mathbb C_{\pm 1}} = \sigma(h)_{\mathbb C_{\pm 1}}$ with $\sigma(h)_{\mathbb C_{\pm 1}} = \rho(h)_{\mathbb C_{\pm 1}}^{-1}$. On the other hand, the naturality condition is always satisfied for $\mathbb C_{-1}\otimes -$. It follows that there is a unique simple 1-intertwiner $\mc E_{\omega / \bar \omega} \to \mc E_{\omega / \bar \omega}$, which we notate via $n_{\omega / \bar \omega}$.

\medskip \noindent
$\mc E_{1} \to \mathbb 1$: This scenario is different than the previous ones in that the $\Vect_{\Zt}$-module categories underlying the 2-representations $\mc E_1$ and $\mathbb 1$ differ so that such a 1-interwiner is now specified by a choice of objects in $\Fun_{\Vect_{\Zt}}(\Vect_{\Zt},\Vect)$. But the category $\Fun_{\Vect_{\Zt}}(\Vect_{\Zt},\Vect)$ is equivalent to $\Vect$, whereby the unique object in $\Vect$ amounts to the forgetful functor $\Vect_{\Zt} \to \Vect$. The naturality condition further stipulates that $\rho(h)_{\mathbb C_{\pm 1}} = \sigma(h)_{\mathbb C}$ for every $h \in \ZT$, which is always satisfied since both $\mc E_1$ and $\mathbb 1$ are associated with the trivial character of $\mathbb Z_3$. We denote this unique 1-intertwiner $\mc E_1 \to \mathbb 1$ by $\rdw$. By transposition, we find a unique 1-intertwiner $\mathbb 1 \to \mc E_1$, which we denote by $\ldw$. These provide the domain walls between the condensation defect and the vacuum found in the previous section. 

\bigskip \noindent
It follows immediately from the previous derivations that there are no 1-intertwiners of the form $\mc E_{1} \to \mc E_{\omega/\bar \omega}$, $\mathbb 1 \to \mc E_{\omega / \bar \omega}$, or transpositions thereof, as trivial and non-trivial characters of $\ZT$ are incompatible preventing the naturality condition \eqref{eq:NatIntb} from being satisfied. This is consistent with the absence of domain walls between the corresponding 1-charges witnessed in the previous section and concludes our analysis of the 1-intertwiners.

\subsection{Composition rule}

\noindent
We established above that the domain walls derived in the previous section are provided by the 1-intertwiners in $\TRep(\mathbb G(2,3))$. We shall now confirm that the fusion of these domain walls in encoded into the composition rule of the corresponding 1-intertwiners. Generally speaking, given two 1-intertwiners $\vartheta : \rho \to \sigma$ and $\tilde \vartheta : \sigma \to \varsigma$, their horizontal composition is defined as the 1-intertwiner $\vartheta \circ \tilde \vartheta$ such that
\begin{equation}
    \label{eq:compOneInt}
    \begin{split}
    (\vartheta \circ \tilde \vartheta)_{\bulScr} &:= \vartheta_{\bulScr} \circ \tilde \vartheta_{\bulScr} \in \Fun(\rho(\bul),\varsigma(\bul))
    \\
    (\vartheta \circ \tilde \vartheta)_g &:= 
    (\vartheta_g \circ 1_{\tilde \vartheta_{\bulScr}}) \cdot (1_{\vartheta_{\bulScr}} \circ \tilde \vartheta_g) 
    \end{split} \, ,
\end{equation}
for every $g \in G$. This simply expresses the composition of the $\Vect_{\Zt}$-module functors underlying the 1-intertwiners. In particular, the defining isomorphism associated with the module functor $(\vartheta \circ \tilde \vartheta)_{\bulScr}$ is given by $((\vartheta \circ \tilde \vartheta)_g)_M := \tilde \vartheta_{\bulScr}(\omega_{M,\mathbb C_g}) \circ \tilde \omega_{\vartheta_{\bulScr}(M),\mathbb C_g}$ for any $M \in \Ob(\rho(\bul))$ and $g \in \Zt$.

\bigskip \noindent
$*$ We established above that simple 1-intertwiners $\mathbb 1 \to \mathbb 1$ are in one-to-one correspondence with simple objects in $\Fun_{\Vect_{\Zt}}(\Vect,\Vect) \cong \Rep(\Zt)$. It follows from the definition of $((\vartheta \circ \tilde \vartheta)_g)_M$ given above that (horizontal) composition of these 1-intertwiners is provided by the monoidal structure in $\Rep(\Zt)$, and thus we have $e \circ e \simeq 1_{\mathbb 1}$. Similarly composition of 1-intertwiners of the form $\mc E_1 \to \mc E_1$ is provided by the monoidal structure in $\Vect_{\Zt}$ and thus $n \circ n \simeq 1_{\mc E_1}$. These composition rules are compatible with the fusion rules of the corresponding domain walls derived in the previous section.

\medskip \noindent
$*$ Analogously to the previous cases, the 1-intertwiners $\rdw : \mc E_1 \to \mathbb 1$ and $\ldw : \mathbb 1 \to \mc E_1$ are provided by the unique simple objects in $\Fun_{\Vect_{\Zt}}(\Vect_{\Zt},\Vect) \cong \Vect$ and $\Fun_{\Vect_{\Zt}}(\Vect,\Vect_{\Zt}) \cong \Vect$, respectively. Compositions $\ldw \circ \rdw$ and $\rdw \circ \ldw$ of the 1-intertwiners are then dictated by the compositions of the corresponding $\Vect_{\Zt}$ module functors \cite{2002math......2130O}. By definition $\ldw \circ \rdw$ is an object in $\Fun_{\Vect_{\Zt}}(\Vect,\Vect) \cong \Rep(\Zt)$ and can be shown to correspond to the \emph{regular} representation of $\mathbb Z_2$. It follows that $\ldw \circ \rdw \simeq 1_\mathbb 1 \oplus e$. Similarly, we find that $\rdw \circ \ldw \circ 1_{\mc E_1} \oplus n$ as an object in $\Fun_{\Vect_{\Zt}}(\Vect_{\Zt},\Vect_{\Zt}) \cong \Vect_{\Zt}$. 

\medskip \noindent
$*$ Let us finally consider the composite $n_\omega \circ n_{\bar \omega}$. Since the composition of 1-intertwiners boils down to the composition of the corresponding module functors, we already know that we must have $n_\omega \circ n_{\bar \omega} \simeq 1_{\mc E_\omega}$ as it is the unique 1-intertwiner $\mc E_{\omega} \to \mc E_{\omega}$, but let us check explicitly how the composition behaves with respect to the naturality condition \eqref{eq:NatIntb}. Recall that both $n_\omega$ and $n_{\bar \omega}$ corresponds to $\mathbb C_{-1} \otimes -$ as the unique simple $\Vect_{\Zt}$-module endofunctor of $\Vect_{\Zt}$ satisfying the naturality condition. It follows from eq.~\eqref{eq:compOneInt} that $n_\omega \circ n_{\bar \omega}$ corresponds to $\mathbb C_{+1} \otimes -$ and that the naturality condition for 1-intertwiners $\mc E_{\omega} \to \mc E_{\omega}$ is satisfied as $\rho(h)_{\mathbb C_{\pm 1}} = \rho(h)_{\mathbb C_{-1} \otimes (\mathbb C_{-1} \otimes \mathbb C_{\pm 1})}$ for every $h \in \mathbb Z_3$.

\subsection{Monoidal structure}

\noindent
We close our dictionary between boundary excitations of the higher gauge model and the bicategory of 2-representations of $\mathbb G(2,3)$ with the recovery of the fusion of the 1-charges from the monoidal structure in $\TRep(\mathbb G(2,3))$. 

Recall that the simple objects in $\TRep(\mathbb G(2,3))$ associated with the 1-charges $\mc E_1$, $\mc E_\omega$ and $\mc E_{\bar \omega}$ are specified by the $\Vect_{\Zt}$-module category $\Vect_{\Zt}$ and the characters of $\ZT$ labelled by $1$, $\omega$ and $\bar \omega$, respectively. A (right) $\Vect_{\Zt}$-module structure can be readily defined on the product $\Vect_{\Zt} \boxtimes \Vect_{\Zt}$ via $(M \boxtimes N) \cat \mathbb C_g := (M \cat \mathbb C_g) \boxtimes (N \cat \mathbb C_g) = (M \otimes \mathbb C_g) \boxtimes (N \otimes \mathbb C_g)$ for any $M,N \in \Ob(\Vect_{\Zt})$ and $g \in G$ \cite{Greenough_2010}. The resulting module category decomposes as $\Vect_{\Zt} \boxplus \Vect_{\Zt}$, where the first copy of $\Vect_{\Zt}$ has simple objects $\mathbb C_{\pm 1} \boxtimes \mathbb C_{\pm 1}$ and the latter has simple objects $\mathbb C_{\pm 1} \boxtimes \mathbb C_{\mp 1}$. This already provides the fusion rule $\mc E_{1} \boxtimes \mc E_1 \cong \mc E_1 \boxplus \mc E_1$. It remains to understand how this tensor structure interacts with non-trivial characters of $\ZT$.

Consider the 2-representations $\rho \equiv \mc E_{\eta}$ and $\sigma \equiv \mc E_{\mu}$, where $\eta,\mu \in \{1,\omega,\bar \omega\}$ label characters of $\ZT$. Following the derivation above, we are looking for 2-representations $\varsigma_1$ and $\varsigma_2$ such that $\rho \boxtimes \sigma \cong \varsigma_1 \boxplus \varsigma_2$ with $\varsigma_1(\bul) = \varsigma_2(\bul) = \Vect_{\Zt}$ as $\Vect_{\Zt}$-module categories. Moreover, we established that simple objects in $\varsigma_1(\bul)$ are given by $\mathbb C_{\pm 1} \boxtimes \mathbb C_{\pm 1}$, whereas simple objects in $\varsigma_2(\bul)$ are given by $\mathbb C_{\pm 1} \boxtimes \mathbb C_{\mp 1}$. The representations of $\mathbb Z_3$ associated with $\varsigma_1$ and $\varsigma_2$ are then defined via $\varsigma_1(-)_{\mathbb C_{\pm 1} \boxtimes \, \mathbb C_{\pm 1}} := \rho(-)_{\mathbb C_{\pm 1}} \otimes \sigma(-)_{\mathbb C_{\pm 1}}$ and $\varsigma_2(-)_{\mathbb C_{\pm 1} \boxtimes \, \mathbb C_{\mp 1}} := \rho(-)_{\mathbb C_{\pm 1}} \otimes \sigma(-)_{\mathbb C_{\mp 1}}$, respectively. Let us check that these representations are well-defined. In particular, they must be compatible with the action of $\Vect_{\Zt}$ on  $\varsigma_1(\bul)$ and $\varsigma_2(\bul)$.
Recall that we have $\rho(g \act - )_{\mathbb C_{\pm 1}} = \rho(-)_{\mathbb C_{\pm g}}$ and $\sigma(g \act - )_{\mathbb C_{\pm 1}} = \sigma(-)_{\mathbb C_{\pm g}}$ for every $g \in G$. It follows that we have
\begin{align}
    \varsigma_{1}(g \act -)_{\mathbb C_{\pm 1} \boxtimes \, \mathbb C_{\pm 1}}
    &= \rho(g \act -)_{\mathbb C_{\pm 1}} \otimes \sigma(g \act -)_{\mathbb C_{\pm 1}}
    \\ \nn
    &= \rho(-)_{\mathbb C_{\pm 1} \otimes \,  \mathbb C_g} \otimes \sigma(-)_{\mathbb C_{\pm 1} \otimes \, \mathbb C_g}
    \\ \nn
    &= \varsigma_1(-)_{(\mathbb C_{\pm 1} \otimes \, \mathbb C_g) \, \boxtimes \, (\mathbb C_{\pm 1} \otimes \, \mathbb C_g)}
    \\ \nn
    &= \varsigma_1(-)_{(\mathbb C_{\pm 1} \boxtimes \, \mathbb C_{\pm 1}) \cat \mathbb C_g} 
\end{align}
for every $g \in G$, as requested, and similarly for $\varsigma_2$. Recall finally that while $\varsigma_1(-)_{\mathbb C_{+1}}$ corresponds to the character $\eta$, $\varsigma_1(-)_{\mathbb C_{-1}}$ corresponds to $\bar \eta$, and similarly for $\varsigma_2$ with respect to $\mu$. Putting everything together, we recover the fusion rules $\mc E_{\eta} \boxtimes \mc E_{\mu} \cong \mc E_{\eta \mu} \boxplus \mc E_{\eta \bar \mu}$, for any $\eta,\mu \in \{1,\omega,\bar \omega\}$. This concludes our analysis of the bicategory $\TRep(\mathbb G(2,3))$.

\section{Discussion}

\noindent
Relying on the paradigm that global symmetry operators in any quantum model correspond to topological defects, we explored in this manuscript the (categorical) charges associated with a simple case of higher group symmetry. More precisely, we studied the excitation content of the symmetry-preserving boundary of a higher gauge topological model mixing 0-form $\mathbb Z_2$- and 1-form $\mathbb Z_3$-gauge symmetries.

Partially guided by the physical interpretation of the non-trivial 2-representation of the (1-)group $\mathbb Z_2$ as a string-like condensation defect in the (3+1)d toric code \cite{PhysRevB.96.045136,Kong:2020wmn}, we proposed an interpretation of the simple 2-representations of the 2-group $\mathbb G(2,3)$ as composites of condensation defects and violations of 1-form gauge invariance in the higher gauge model. Although we focused on $\mathbb G(2,3)$ for concreteness, we could have dealt with the dihedral 2-group $\mathbb G(2,n)$ for an arbitrary $n$ with no additional effort, and with the more general case discussed in sec.~\ref{sec:TRep} almost as easily. We expect the case of an arbitrary 2-group to be treatable invoking similar techniques, however the interpretation we offered of 2-representations and 1-intertwiners in terms of module categories and module category functors would have to be adapted. That being said, we expect the interplay between condensation defects---gases of 0-charges that spontaneously break the effective 0-form symmetry along one-dimensional submanifolds---and 1-charges to persist in the most general case.

\bigskip\bigskip\noindent
{\bf Acknowledgements:} The author is grateful to Joe Huxford for answering questions about manuscript \cite{Huxford:2022wih}, as well as Alex Bullivant and Apoorv Tiwari for numerous discussions on closely related topics.

\bigskip \noindent
{\bf Note added:} After completion of this work, the author was made aware of the manuscripts \cite{Bartsch:2022mpm} and \cite{Bhardwaj:2022lsg} that might contain partially overlapping results.

\bigskip \bigskip
\bibliography{ref}	

\end{document}